\documentclass[letterpaper, oneside, 12pt]{article}

\usepackage{amsmath}
\usepackage{graphicx}%
\usepackage{amsfonts}%
\usepackage{amssymb}
\usepackage{xcolor}
\usepackage{overpic}
\usepackage[ruled]{algorithm}
\usepackage[noend]{algpseudocode}
\usepackage{subfigure}
\usepackage{rotating}
\usepackage{epstopdf}
\usepackage{url}
\usepackage{comment}
\usepackage{enumerate}
\usepackage{mathtools}
\usepackage{multirow}
\usepackage{pgfplots}
\usepackage[numbers, square]{natbib}

\usepackage{pst-solides3d}
\usepackage[section]{placeins}  % To make the figure appear before the appendix.
\usepackage{tikz}
\usetikzlibrary{arrows.meta, positioning, shapes, fit}

\usepackage{setspace}
\usepackage[hmargin=1.0in,vmargin=1.0in]{geometry}

\usepackage[compact]{titlesec}
\titlespacing{\section}{0pt}{*0.8}{*0.8}
\titlespacing{\subsection}{0pt}{*0.8}{*0.8}
\titlespacing{\subsubsection}{0pt}{*0.8}{*0.8}

\newcommand{\ba}{ {\boldsymbol a} }
\newcommand{\bA}{ {\boldsymbol A} }

\newcommand{\bB}{ {\boldsymbol B} }

\newcommand{\bI}{ {\boldsymbol I} }

\newcommand{\bJ}{ {\boldsymbol J} }
\newcommand{\bk}{ {\boldsymbol k} }

\newcommand{\bO}{ {\boldsymbol O} }

\newcommand{\bx}{ {\boldsymbol x} }

\newcommand{\by}{ {\boldsymbol y} }
\newcommand{\bY}{ {\boldsymbol Y} }

\newcommand{\bbeta}{ {\boldsymbol \beta} }

\newcommand{\bgamma}{ {\boldsymbol \gamma} }
\newcommand{\bGamma}{ {\boldsymbol \Gamma} }

\newcommand{\bLambda}{ {\boldsymbol \Lambda} }
\newcommand{\bmu}{ {\boldsymbol \mu} }

\newcommand{\bSigma}{ {\boldsymbol \Sigma} }

\newcommand{\bzero}{ {\boldsymbol 0} }

\usepackage{titlesec}

\titleformat{\paragraph}
{\normalfont\normalsize\bfseries}{\theparagraph}{1em}{}
\titlespacing*{\paragraph}
{0pt}{3.25ex plus 1ex minus .2ex}{1.5ex plus .2ex}

\newtheorem{theorem}{Theorem}[section]

\newcommand{\qed}{\nobreak \ifvmode \relax \else
      \ifdim\lastskip<1.5em \hskip-\lastskip
      \hskip1.5em plus0em minus0.5em \fi \nobreak
      \vrule height0.75em width0.5em depth0.25em\fi}
\title{Multiview Graph Fusion with Covariates}

\author{Sharmistha Guha\footnotemark[1]
\and
Jose Rodriguez-Acosta\footnotemark[1]
    \and 
Ivo Dinov \footnotemark[2]
}

\date{}

\begin{document}

\maketitle

\footnotetext[1]{Department of Statistics, Texas A\&M University}

\footnotetext[2]{Statistics Online Computational Resource, Computational Medicine \& Bioinformatics, University of Michigan}

\begin{abstract}
Joint modeling of multiview graphs with a common set of nodes between views and auxiliary predictors is an essential, yet less explored, area in statistical methodology. Traditional approaches often treat graphs in different views as independent or fail to adequately incorporate predictors, potentially missing complex dependencies within and across graph views and leading to reduced inferential accuracy. Motivated by such  methodological shortcomings, we introduce an integrative Bayesian approach for joint learning of a multiview graph with vector-valued predictors. 
Our modeling framework assumes a common set of nodes for each graph view while allowing for diverse interconnections or edge weights between nodes across graph views, accommodating both binary and continuous valued edge weights. By adopting a hierarchical Bayesian modeling approach, our framework seamlessly integrates information from diverse graphs through carefully designed prior distributions on model parameters. This approach enables the estimation of crucial model parameters defining the relationship between these graph views and predictors, as well as offers predictive inference of the graph views. Crucially, the approach provides uncertainty quantification (UQ) in all such inferences.
Theoretical analysis establishes that the posterior predictive density for our model asymptotically converges to the true data-generating density, under mild assumptions on the true data-generating density and the growth of the number of graph nodes relative to the sample size. Simulation studies validate the inferential advantages of our approach over predictor-dependent tensor learning and independent learning of different graph views with predictors. We further illustrate model utility by analyzing functional connectivity (FC) graphs in neuroscience under cognitive control tasks, relating task-related brain connectivity with phenotypic measures.
\end{abstract}

\noindent{\emph{Keywords:}} Multiview graph response,
Posterior consistency,
Hierarchical Bayesian modeling,
High-dimensional regression,
Functional connectivity

\newpage

\section{Introduction}
The analysis of multiview relational data structures, often represented as graphs or networks, has become increasingly important in modern statistical methodology. In various scientific domains, researchers encounter settings in which multiview graphs, defined over a common set of entities or nodes, are observed for each subject or unit. These multiview graphs can reflect interactions between nodes under different conditions or domains and are often accompanied by subject-level predictor information. Joint predictor-dependent learning of different views of a multiview graph offers a principled approach to uncover associations between graph structures and subject-level predictors.

\noindent \textbf{Existing literature on predictor-dependent joint learning of graphs.} In the domain of predictor-dependent joint learning for multiview graphs, a commonly employed strategy involves deriving various summary metrics from each graph. Subsequently, a regression framework is applied to establish joint associations between these summary metrics from different graphs and the predictors \citep{bullmore2009complex}. However, a significant drawback of this approach is its susceptibility to the choice of summary metrics, which can profoundly affect the inference. To mitigate this limitation, some recommendations advocate transforming each graph into a high-dimensional vector of edge weights. A regression model is then fitted on these vectors using the predictors, resulting in a high-dimensional vector-on-scalar regression problem. This strategy can leverage recent advancements in high-dimensional multivariate reduced-rank regression \citep{rothman2010sparse, chen2012sparse, goh2017bayesian}. Nevertheless, adopting this approach in our context presents two major drawbacks. First, this method overlooks potential correlations among coefficients associated with edges that share common graph nodes. Second, our inferential objective involves modeling different graph views with both binary and continuous edge weights. To our knowledge, no existing literature addresses the modeling of binary and continuous high-dimensional responses simultaneously within a multivariate reduced-rank regression framework. 

Another potential approach is to represent the multiview graph as a tensor and leverage established methods for predictor-dependent tensor learning, which can incorporate various combinations of low-rank and sparsity assumptions on the tensor-valued predictor coefficients \citep{rabusseau2016low, spencer2020joint, lee2023bayesian, guhaniyogi2021bayesian}. However, these methods typically do not explicitly enforce the symmetry constraint on the predictor coefficients associated with each graph view. Given that each view corresponds to an undirected graph, such a symmetry constraint is desirable; omitting it may result in predictor coefficients that do not fully capture the inherent structure of undirected graphs, making scientific interpretation more challenging.
In this context, \cite{guha2021bayesiantensor} and \cite{guha2024covariate} proposed a graph-on-predictors regression framework that incorporates symmetry constraints on the predictor coefficients. Their approach involves specifying a prior distribution that imposes sparsity on parameters capturing the effects of graph nodes and edges in the regression between multiview graphs and predictors. However, their methodology is restricted to modeling the relationship between a single graph and its predictors. To the best of our knowledge, there are currently no methods available for joint predictor-dependent learning of multiview graphs that also provide uncertainty quantification for both model parameters and predictions. Recent work on scalar-on-multiview graph regression \citep{rodriguez2025multiplex, guha2024bayesian} also diverges from our goals, as these approaches use the multiview graph solely as a predictor without jointly modeling the graph structure and other variables of interest. As a result, they do not facilitate inference regarding differences in graph structure, for example, across various cognitive control domains for the study of functional connectivity (see Section~\ref{sec:real_data_analysis}), which may underlie the observed variation among graph views.

Our methodological problem fundamentally differs from the literature on the joint estimation of multiple graphs, which is commonly employed in the study of genomic and neuroimaging data. In this literature, the focus is on jointly estimating multiple \emph{unknown} graph structures from multivariate observations obtained over a set of nodes using multiple Gaussian graphical models (GGMs) or their variants, with or without accounting for subject-specific predictors \citep{guo2011joint, danaher2014joint, peterson2015bayesian, lukemire2021bayesian, coombes2015weighted, liu2010graph, cheng2014sparse, niu2023covariate}. In contrast, our framework assumes multiview graphs as observed data, with the primary focus on predictor-dependent modeling of these graphs.

Our approach diverges from existing literature on the joint modeling of graphs \citep{wang2013exponential,gollini2016joint} in several critical aspects. First, the existing literature is largely unsupervised, indicating that the joint modeling of graphs is conducted without incorporating subject-level predictors. Second, the primary objective of this literature revolves around drawing inference on the dependence between  multiple graphs across nodes. In contrast, our inferential focus is distinct and is centered on making inferences regarding the relationship between each graph and subject-level predictors of interest, jointly learned from multiple views. In fact, the emphasis of this article is not on statistical testing or inference concerning the association between heterogeneous graphs.

Graph neural networks (GNNs) have introduced flexible architectures for learning representations from multiview graphs conditioned on predictors. Multiview GNNs fuse information across views using attention, hierarchical aggregation, or neural fusion techniques \citep{sun2022a2ae, wei2023multi,shen2022semi}. Recent extensions include joint modeling of node/edge features and external predictors, allowing predictions or inferences conditioned on observed attributes \citep{xiao2024graph}. These methods often support end-to-end learning, but embedding interpretability and formal statistical inference remain ongoing challenges.

\noindent \textbf{Outline of the proposed approach.} This article presents a generalized linear modeling approach aimed at investigating the joint relationship between graph views, including both binary and continuous edge weights, and scalar predictors. The \emph{graph coefficients}, representing the relationship between each graph and each scalar predictor, incorporate a low-rank structure involving latent vectors representing effects of nodes from each graph in determining their relationships with a predictor. This assumed low-rank structure significantly reduces the number of parameters required to estimate graph coefficients and enforces desirable graph properties, such as transitivity \citep{guha2021bayesiantensor,guha2024covariate}, in the modeling of graphs. To facilitate joint estimation and inferences of model parameters, we adopt a hierarchical Bayesian framework, leveraging its ability to integrate information across graphs through carefully structured joint prior distributions on graph coefficients. Specifically, we assign a spike-and-slab variable selection prior to the latent vectors associated with a node across all graphs jointly. This approach enables direct inference on nodes and edges that exhibit a significant relationship with a predictor, jointly estimated from all views. The proposed framework enables efficient Bayesian computation, as well as uncertainty quantification (UQ) in inference.

Additionally, we explore the theoretical properties of our proposed approach, demonstrating that the predictive density of the proposed generalized linear model for diverse graph views converges to the true data-generating model. These theoretical results provide valuable insights into how the number of nodes in multiview graphs, the dimensions of node-specific latent vectors, and the structure and sparsity of true coefficients corresponding to a predictor can adapt with changing sample sizes (denoted as ``n'') to achieve asymptotically accurate estimates of the true data-generating predictive density.

\noindent\textbf{Novelty of the proposed approach.} \textbf{(1) Joint learning of multiview graph adjusted for predictors.}  Our framework imposes predictor-dependent association between multiple graph views, draws inference on graph nodes influentially related to each predictor and offers inference on coefficients for graph edges that encapsulate relationships between different graphs and predictors. Moreover, the full Bayesian framework allows UQ in inference. To our knowledge, all these inferential goals have not been simultaneously achieved before. 
\textbf{(2) Theoretical results.} Our theoretical exposition introduces several novel aspects over existing work in Bayesian predictor-dependent learning of multiple graphs. Firstly, the theoretical framework in this article addresses joint modeling with multiview graph responses, unlike scenarios of a single graph response addressed in prior literature \citep{guha2023high,guha2021bayesiantensor,guha2023covariate}. It is also important to emphasize that our results are distinct from posterior contraction properties in high-dimensional predictor-dependent learning of tensor objects \citep{guhaniyogi2017bayesian, guha2021bayesiantensor, guhaniyogi2021bayesian}. In particular, these results do not naturally extend to our framework, as they do not incorporate symmetry in the coefficient tensors corresponding to each predictor, a requirement in our framework involving a symmetric graph matrix for each view. Second, the theoretical framework accommodates different graph views modeled with distinct link functions, allowing for the theoretical umbrella to include continuous, binary, or categorical edge weights in various graph views. Moreover, the true graph coefficients, which capture associations between different graph views and scalar predictors, are assumed to exhibit low-rank structures with different ranks corresponding to different graphs. The aforementioned issues pose considerable theoretical challenges beyond what is encountered in \cite{guha2023high}, \cite{guha2021bayesiantensor}, and \cite{guha2023covariate}, necessitating novel proof techniques, as outlined in Appendix A. Our work presents novel Bayesian asymptotic results on predictor-dependent joint learning of heterogeneous graph structures in different views, that, to our knowledge, have not been previously established.
\textbf{(3) Generalizability.} The proposed approach assumes straightforward extension to incorporate  joint modeling of heterogeneous graphs and node-specific predictors,  along with subject-level predictors considered here, extending the approaches in \cite{fosdick2015testing,guhaniyogi2020joint}. 
\textbf{(4) Study of FC in diverse cognitive control domain.} Using multiview task-based fMRI graphs, the proposed joint learning approach identified functional connectivity patterns associated with mini-mental state examination (MMSE) performance that were concentrated in networks supporting executive, attentional, and affective–mnemonic processes, including the dorsal attention, control, salience/ventral attention, and limbic systems. This provides a novel, principled framework for linking network-level connectivity during cognitive control tasks to cognitive aging.

The rest of the article proceeds as follows. Section~\ref{sec:description} provides details of the model development and prior distributions on the model coefficients. Section~\ref{sec:post_conv} details the posterior contraction result for the predictive density and the sufficient conditions for the result to hold. Section~\ref{sec:post_comp} discusses posterior computation. Empirical performance of the proposed approach is demonstrated through simulation studies in Section~\ref{sec:simulation}. Section~\ref{sec:real_data_analysis} describes the scientific problem on the study of functional connectivity across diverse tasks of cognitive control and employs our approach to analyze this data. Finally, Section~\ref{sec:conclusions} summarizes the contribution and discusses scope for future work. Proofs of the theoretical results are presented in Appendix A, while Appendix B shows full conditional distributions to construct a Gibbs sampler for parameter estimation.

\section{Predictor-Dependent Joint Learning of Heterogeneous Graphs}\label{sec:description}
\subsection{Notations}
 Throughout the article, we denote a scalar by a lower or upper case letter, for example, $a$ or $A$; a vector by a boldface lowercase letter and a matrix by a boldface uppercase letter, $\ba$ and $\bA$, respectively. 
For $i=1,...,n$, let $\mathcal{G}_{i,1},...,\mathcal{G}_{i,M}\in\mathcal{Y}$ denote the weighted undirected $M$ graph views of the multiview graph defined on a common set of $K$ nodes denoted by $\mathcal{N}=\{\mathcal{N}_1,...,\mathcal{N}_K\}$. In our application of interest in Section~\ref{sec:real_data_analysis}, these labelled nodes correspond to the regions of interest in a human brain. The undirected graph $\mathcal{G}_{i,m}$ can be represented by a symmetric $K\times K$ matrix $\bY_{i,m}$, $m=1,...,M$, with its $(k,k')$th entry $y_{i,m,(k,k')}$  satisfying $y_{i,m,(k,k')}=y_{i,m,(k',k)}$. Since our motivating application in Section~\ref{sec:real_data_analysis} does not attach any significance to self-relationship in nodes of the graph, we assume 
$y_{i,m,(k,k)}=0$, for all $k=1,...,K$, without loss of generality. 
While all $M$ graphs are defined on the same set of labelled nodes, we allow the edges in different graphs to be continuous or binary. Let 
$\mathcal{M}_c$ and $\mathcal{M}_b$ denote sets of indices corresponding to graphs with continuous and binary edges, respectively.
 Thus, $y_{i,m,(k,k')}\in\{0,1\}$ if $m\in\mathcal{M}_b$ and $y_{i,m,(k,k')}\in\mathbb{R}$ if $m\in\mathcal{M}_c$, with $\mathcal{M}_c\cup\mathcal{M}_b=\{1,...,M\}$ and $\mathcal{M}_c\cap\mathcal{M}_b=\emptyset$. %Thus, $\bY_{i,m}$, for $m\in\mathcal{M}_b$ and $m\in\mathcal{M}_c$, represent binary and continuous weighted networks, respectively.

For subject $i$, $i=1,..,n$, our proposed model makes an assumption concerning two distinct sets of predictors. The first set of predictors, denoted as $\bx_i=(x_{i,1},...,x_{i,P})^T$, is assumed to have varying relationships with the elements $y_{i,m,(k,k')}$ across different cells of the graph matrix. In contrast, the second set of predictors, represented as $\tilde{\bx}_i=(\tilde{x}_{i,1},...,\tilde{x}_{i,\tilde{P}})^T$, is presumed to exert a uniform influence on every cell of the graph matrix. These two sets of predictors are referred to as ``key predictors'' and ``auxiliary predictors,'' respectively. In the context of Section~\ref{sec:real_data_analysis}, the cognitive score from the Mini-Mental State examination serves as a ``key predictor,'' while age and gender are ``auxiliary predictors.''
 Let $\mathcal{K}=\{\bk=(k_1,k_2): 1\leq k_1<k_2\leq K\}$ be a set of indices. Since every $\bY_{i,m}$ is symmetric with $0$ diagonal entries, it suffices to build a probabilistic generative mechanism for every entry $y_{i,m,\bk}$ $(\bk\in\mathcal{K})$ in the upper triangle, as elaborated in the subsequent sections.
 
\subsection{Model Development}\label{sec:model}
For subject $i=1,...,n$, we propose a set of conditionally independent generalized linear models to build a regression relationship between graph views $\bY_{i,1},...,\bY_{i,M}$ and predictors, given by
\begin{align}\label{eq:joint_network_model}
E[y_{i,m,\bk}] = u_{i,m,\bk}=G_m^{-1}(\mu_m+\sum_{p=1}^Px_{i,p}\gamma_{p,m,\bk}+\sum_{\tilde{p}=1}^{\tilde{P}}\tilde{x}_{i,\tilde{p}}\alpha_{
\tilde{p},m}),\:\:\bk\in\mathcal{K},
\end{align}
where $G_m(\cdot)$ is the link function corresponding to the edges in the $m$th graph and $\gamma_{1,m,\bk}$,...,$\gamma_{P,m,\bk}\in\mathbb{R}$ express the effect of the $P$ key predictors on the $\bk$th cell of the $m$th graph matrix. Furthermore, $\gamma_{1,m,\bk}$,...,$\gamma_{P,m,\bk}$ are assumed to be the $\bk$th cell entries of the symmetric matrices
$\bGamma_{1,m},...,\bGamma_{P,m}\in\mathbb{R}^{K\times K}$, respectively, each with zero diagonal entries. The coefficients for the auxiliary predictors are given by $\alpha_{1,m},...,\alpha_{\tilde{P},m}\in\mathbb{R}$, and $\mu_m\in\mathbb{R}$ represents the regression intercept. While all $M$ graphs are regressed on the same set of predictors, equation (\ref{eq:joint_network_model}) ensures varying regression effects of a predictor on different graphs by adding graph-specific coefficients for predictors. Using the identity link function for graphs having continuous edge effects and the logit link function for graphs with binary edge effects, equation (\ref{eq:joint_network_model}) becomes
\begin{align}\label{eq:cont_and_binary}
 y_{i,m,\bk} &= \mu_m+\sum_{p=1}^Px_{i,p}\gamma_{p,m,\bk}+\sum_{\tilde{p}=1}^{\tilde{P}}\tilde{x}_{i,\tilde{p}}\alpha_{
\tilde{p},m}+\epsilon_{i,m,\bk},\:\:\mbox{for}\:m\in\mathcal{M}_c\nonumber\\
P(y_{i,m,\bk}=1)&=\frac{\exp\left(\mu_m+\sum_{p=1}^Px_{i,p}\gamma_{p,m,\bk}+\sum_{\tilde{p}=1}^{\tilde{P}}\tilde{x}_{i,\tilde{p}}\alpha_{
\tilde{p},m}\right)}{1+\exp\left(\mu_m+\sum_{p=1}^Px_{i,p}\gamma_{p,m,\bk}+\sum_{\tilde{p}=1}^{\tilde{P}}\tilde{x}_{i,\tilde{p}}\alpha_{
\tilde{p},m}\right)},\:\:\mbox{for}\:m\in\mathcal{M}_b.
\end{align}
For $m\in\mathcal{M}_c$, $\epsilon_{i,m,\bk}$ are the idiosyncratic errors following i.i.d. N($0,\sigma_m^2$).

To enhance parsimony in the estimation of high-dimensional symmetric matrices $\bGamma_{1,m},\ldots,\bGamma_{P,m}$, we introduce a low-rank structure to these matrices given by
\begin{align}\label{eq:low_rank}
\gamma_{p,m,\bk}=\sum_{r=1}^{R}\lambda_{p,m}^{(r)}\beta_{p,m,k_1}^{(r)}\beta_{p,m,k_2}^{(r)},\quad m=1,\ldots,M;\quad \bk\in\mathcal{K},
\end{align}
where $\tilde{\bbeta}_{p,m,k}=(\beta_{p,m,k}^{(1)},\ldots,\beta_{p,m,k}^{(R)})^T$, $m=1,\ldots,M$, is a collection of $R$-dimensional latent variables, one for each node, each key predictor, and each graph view. The quantity $\tilde{\bbeta}{p,m,k}$ represents the effect of the $p$th key predictor on the $k$th node of the $m$th graph view. The construction of $\bGamma{p,m}$ induced by equation (\ref{eq:low_rank}) draws inspiration from the low-rank decomposition of matrices. By defining $\tilde{\bB}_{p,m}$ as a $K\times R$ matrix with its $k$th row as $\tilde{\bbeta}_{p,m,k}$ and $\bLambda_{p,m}$ as a $R\times R$ diagonal matrix with its $r$th diagonal entry $\lambda_{p,m}^{(r)}$, equation (\ref{eq:low_rank}) represents a low-rank decomposition of $\bGamma_{p,m}$: $\bGamma_{p,m}=\tilde{\bB}_{p,m}\bLambda_{p,m}\tilde{\bB}_{p,m}^T$. The selection of $R$ is user-defined, with the prior construction on $\lambda_{p,m}^{(r)}$ helping to prevent overfitting and promote model efficiency, as elaborated in subsequent discussions.

The low-rank decomposition presented in equation (\ref{eq:low_rank}) necessitates the estimation of $R(K+1)$ parameters, as opposed to $Q=K(K-1)/2$ parameters needed for estimating each unstructured coefficient matrix. This approach allows for parsimony and efficient computation, especially given that $R<<K$. Additionally, the low-rank structure exhibits a transitivity effect in the model (\cite{guha2021bayesian, guha2023high}). In simpler terms, if the $p$th predictor has no association with the interaction between nodes $k_1$ and $k_2$ (i.e., $\gamma_{p,m,(k_1,k_2)}=0$) and between nodes $k_2$ and $k_3$ (i.e., $\gamma_{p,m,(k_2,k_3)}=0$), then the $p$th predictor is also unrelated to the interaction between nodes $k_1$ and $k_3$ (i.e., $\gamma_{p,m,(k_1,k_3)}=0$). To safeguard the model against overfitting arising from the selection of large values of $R$, the model introduces parameters $\lambda_{p,m}^{(r)}\in\{-1,0,1\}$, which dictate the effect of the $r$th summand in constructing the coefficient $\gamma_{p,m,\bk}$. Specifically, while the fitted dimension of the latent effects is $R$,  $\sum_{r=1}^R|\lambda_{p,m}^{(r)}|\leq R$ determines the data-driven estimation of the dimension of latent variables. 
%referred to as the \emph{intrinsic dimension} of the latent variables.

The identifiability of the node-specific latent variables $\tilde{\bbeta}_{p,m,k}$ depends on the structure of $\bLambda_{p,m}$. For example, if $\bLambda_{p,m} = \bI_R$, then we have 
$\bGamma_{p,m} = \tilde{\bB}_{p,m} \bLambda_{p,m} \tilde{\bB}_{p,m}^T = (\tilde{\bB}_{p,m} \bO) \bLambda_{p,m} (\tilde{\bB}_{p,m} \bO)^T$ for any orthogonal matrix $\bO$. In such cases, unless additional constraints are imposed on $\tilde{\bbeta}_{p,m,k}$, these latent variables may not be identifiable, making direct posterior inference on them potentially uninformative.
Nevertheless, in practice, our primary inferential focus is not the latent node-specific variables themselves. Inference on edge coefficients is based on estimating $\bGamma_{p,m}$, which is identifiable. Similarly, to determine whether a particular node is associated with a predictor, we consider the event $\left\{ k : \tilde{\bbeta}_{p,m,k} = \bzero \text{ for all } m = 1, \ldots, M \right\}.$
Even though $\tilde{\bbeta}_{p,m,k}$ is not individually identifiable, this event is identifiable. This aspect of identifiability is essential for conducting inference on the set of nodes related to the $p$th key predictor, as discussed in the following section.
 
\subsection{Joint Prior on Coefficients}\label{sec:prior}
To simultaneously account for association between graph views and draw inference on influential graph nodes, we adopt a hierarchical Bayesian approach and propose a joint prior distribution on $\bGamma_{p,1},...,\bGamma_{p,M}$. Let
$\tilde{\bbeta}_{p,k}=(\tilde{\bbeta}_{p,1,k}^T,...,\tilde{\bbeta}_{p,M,k}^T)^T$ stack the vector of $k$th node specific latent variable corresponding to the $p$th predictor from all $M$ graphs, $k=1,...,K$, $p=1,...,P$. We assign a spike-and-slab prior jointly on node specific latent variables as below,
\begin{align}\label{prior_u}
\tilde{\bbeta}_{p,k}\sim%\left\{\begin{array}{cc}
\xi_{p,k}N(\bzero, \bJ_{p})+(1-\xi_{p,k})\delta_{\bzero},\:\:
% \end{array}
% \right.,\:\: 
\xi_{p,k}\sim Ber(\eta_{p}),\:\:\bJ_{p} \sim IW(\nu,\bI),\:\:\eta_{p} \sim Beta(1,b_{\eta}),
\end{align}
where $\bJ_{p}$ represents a covariance matrix of size $RM \times RM$, capturing the correlation structure among node-specific latent vectors across all graph views. Specifically, $var(\tilde{\bbeta}_{p,k})=\eta_p\bJ_p$, where $\bJ_p$, $p=1,\ldots,P$, plays a crucial role in establishing the interdependence between $M$ graph views. The indicator variable $\xi_{p,k}$, shared across all graphs, determines the impact of the $p$th predictor on the $k$th graph node. Notably, $\xi_{p,k}=0$ signifies $\tilde{\bbeta}_{p,k}=\bzero$, indicating that the $k$th graph node in all graph views is not associated with the $p$th predictor. Given that edge effects for different graphs can be continuous or binary, the proposed framework accommodates the interdependence between these graph views in the latent space through equation (\ref{prior_u}).

The parameter $\eta_{p}$ corresponds to the probability of the nonzero mixture component in equation (\ref{prior_u}) and is assigned a beta prior to allow multiplicity correction in the Bayesian variable selection framework. The parameters $\lambda_{p,m}^{(r)}$ are assigned a discrete prior distribution, with
$\lambda_{p,m}^{(r)}$ taking values $0,1,-1$ with probabilities $\pi_{p,m,1}^{(r)},\pi_{p,m,2}^{(r)}$ and $\pi_{p,m,3}^{(r)}$, respectively. We set a Dirichlet prior on the probabilities jointly, such that $(\pi_{p,m,1}^{(r)},\pi_{p,m,2}^{(r)},\pi_{p,m,3}^{(r)})\sim Dirichlet(r^{\omega},1,1),\:\:\:  \omega> 1.$
The choice of hyper-parameters of the Dirichlet distribution is crucial.
In particular, $E[|\lambda_{p,m}^{(r)}|]=2/(2+r^{\omega})\rightarrow 0$ as $r\rightarrow\infty$ provides (weak) identifiability of the different latent dimensions and
$\sum_{r=1}^{R}Var(|\lambda_{p,m}^{(r)}|)=
\sum_{r=1}^{R}[\frac{2(r^{\omega}+1)}{(r^{\omega}+2)^2(r^{\omega}+3)}+\frac{2(r^{\omega}+1)}{(r^{\omega}+3)(r^{\omega}+4)}]<\infty$ as $R\rightarrow\infty$ ensures that $\lim_{R\to \infty }Var(\sum_{r=1}^{R}|\lambda_{p,m}^{(r)}|)\leq \lim_{R\to \infty }\sum_{r=1}^{R}Var(|\lambda_{p,m}^{(r)}|<\infty$, since the covariance terms are all negative. This ensures that even if the choice of the fitted dimension $R$ is arbitrarily large, the estimated dimension of the node-specific latent vectors has finite variability a-priori. 
%We assign a hierarchical prior $\lambda_{s,h,r}\sim Ber (\pi_{s,h,r}), \:\pi_{s,h,r} \sim Beta(1, r^{\eta}),\:\eta>1$, 
The parameters $\mu_m,\alpha_{1,m},...,\alpha_{\tilde{P},m}$ are assigned standard normal distributions and the error variance $\sigma_m^2$ is assigned $\textrm{IG}(a_{\sigma},b_{\sigma})$ a-priori.

\section{Posterior Convergence Properties of The Proposed Model}\label{sec:post_conv}
This section presents the convergence properties of the proposed predictor-dependent joint learning framework for multiview graphs. We state and prove two main theoretical results. The first result demonstrates that the predictive density of the proposed joint model converges to the true data-generating density, under the assumption that the fitted models for each graph view belong to the class of generalized linear models (GLMs). Our theoretical framework allows for different GLM densities across the various graph views, thereby accommodating a range of data types such as continuous, binary, or categorical. The second result establishes the convergence of the estimated graph coefficients to the true underlying coefficients. For clarity, this result is presented in the context where all graph edge weights are continuous. We begin by introducing some preliminary notations.

\subsection{Notations}
In our analysis, we add a subscript $n$ to the dimension of the number of graph nodes $K_n$ to indicate an asymptotic setting where the number of graph nodes grows with the sample size. This naturally implies that the $m$th graph view is a function of $n$ and we denote it by $\bY_{n,m}$. Let $\by_{n,m}=\left(y_{n,m,\bk}:\bk=(k_1,k_2)\in\mathcal{K}\right)$ denote the upper triangular part of the graph matrix $\bY_{n,m}$ of dimension $Q_n=K_n(K_n-1)/2$. Given that the graphs are un-directed with no self-relation among nodes, it is enough to focus on the distribution of $\by_{n,m}$. Let $\by_{n}=\left(\by_{n,1}^T,...,\by_{n,M}^T\right)^T$ represent a $MQ_n\times 1$ vector stacking all $\by_{n,m}$'s together. For the sake of algebraic simplicity we assume no auxiliary predictor in the model and the number of key predictors $P=1$; which allows us to drop the subscript $p$ from predictors and coefficients. The true density and predictive density for the $m$th graph $\by_{n,m}$ are assumed to lie in the class of generalized linear models, given by,
\begin{align}\label{general_linear_ model}
   & f_m(\by_{n,m}|x,\bgamma_{n,m})=\prod_{\bk\in\mathcal{K}}f_{m,\bk}(y_{n,m,\bk}|x,\gamma_{n,m,\bk}),\:\:f_m(\by_{n,m}|x,\bgamma_{n,m}^*)=\prod_{\bk\in\mathcal{K}}f_{m,\bk}(y_{n,m,\bk}|x,\gamma_{n,m,\bk}^*)\nonumber\\
 &  f_{m,\bk}(y_{n,m,\bk}|x,\gamma_{n,m,\bk})=\exp(a_m(x\gamma_{n,m,\bk})y_{n,m,\bk}+b_m(x\gamma_{n,m,\bk})+c_m(y_{n,m,\bk}))\nonumber\\
 & f_{m,\bk}(y_{n,m,\bk}|x,\gamma_{n,m,\bk}^*)=\exp(a_m(x\gamma_{n,m,\bk}^*)y_{n,m,\bk}+b_m(x\gamma_{n,m,\bk}^*)+c_m(y_{n,m,\bk})),
\end{align}
where $a_m(w)$ and $b_m(w)$ are continuously differentiable functions, with $a_m(w)$ having a nonzero derivative. This  parameterization includes some popular classes of densities,  e.g., binary logit or probit link and a continuous response with i.i.d.  normal errors having known variance.
In the same spirit as $\by_{n,m}$, we represent the upper triangular vectors of $\bGamma_{n,m}$ and $\bGamma_{n,m}^*$ by $\bgamma_{n,m}$ and $\bgamma_{n,m}^*$, respectively, in equation (\ref{general_linear_ model}). Here $P_{\bgamma_n^*}$ and $P_{\bgamma_n}$ denote probability distributions under the true data generating parameters $\{\bgamma_{n,m}^*:m=1,..,M\}$ and the fitted model parameters $\{\bgamma_{n,m}:m=1,..,M\}$, respectively. In general, $*$ is used to represent the true value of a parameter.

We let $||\cdot||_1, ||\cdot||_{2}$ and $||\cdot||_{\infty}$ denote the $L_1, L_2$ and $L_{\infty}$ norms, respectively. The number of nonzero elements in a vector is given by $||\cdot||_0$. Finally, for two nonnegative sequences $\{d_{1n}\}$ and $\{d_{2n}\}$, we write $d_{1n}\asymp d_{2n}$ to denote $0<\liminf_{n\rightarrow\infty}d_{1n}/d_{2n}\leq \limsup_{n\rightarrow\infty}d_{1n}/d_{2n}<\infty$. If $\lim_{n\rightarrow\infty}d_{1n}/d_{2n}=0$, we write $d_{1n}=o(d_{2n})$ or $d_{1n}\prec d_{2n}$. We use $d_{1n}\lesssim d_{2n}$ or $d_{1n}=O(d_{2n})$ to denote that for sufficiently large $n$, there exists a constant $C>0$ independent of $n$ such that $d_{1n}\leq Cd_{2n}$.
\subsection{Assumptions and Main Result}
We now state the following assumptions on the number of graph nodes and the number of influential nodes.
\begin{enumerate}[(A)]
\item \textbf{Growth of the number of graph nodes and rank of the fitted coefficient:} The number of graph nodes $K_n$ and rank of the fitted graph coefficients $R_n$ grow sub-linearly with the sample size $n$, such that $R_nK_n\prec n/\log(n)$.
\item \textbf{Low-rank decomposition for the true coefficients:} The true coefficients $\bGamma_{n,m}^*$ assume a low-rank decomposition with rank $R_{n,m}^*$, such that $(k_1,k_2)=\bk$th cell of the true coefficient $\bGamma_{n,m}^*$ satisfies $\gamma_{n,m,\bk}^*=\sum_{r=1}^{R_{n,m}^*}\beta_{n,m,k_1}^{*(r)}\beta_{n,m,k_2}^{*(r)}$, for all $\bk\in\mathcal{K}$.
\item \textbf{Growth of the rank of the true coefficient:} The rank $R_{n,m}^*$ of the true coefficient $\bGamma_{n,m}^*$ corresponding to the $m$th graph $\bY_{n,m}$ satisfies $R_{n}\geq R_{n,m}^*$, for all $m=1,...,M$, i.e., the rank of the fitted coefficient must be greater than the rank of the true coefficient.
\item \textbf{Magnitude of cells in the true graph coefficient:} $\beta_{n,m,k}^{*(r)}$ for all $m=1,...,M$; $k=1,..,K_n$ and $r=1,..,R_{n,m}^*$ satisfies $\sum_{m=1}^M\sum_{k=1}^{K_n}||\tilde{\bbeta}_{n,m,k}^{*}||_2\leq \tilde{C}_{\beta}$, for some constant $\tilde{C}_{\beta}>0$, where $\tilde{\bbeta}_{n,m,k}^{*}=(\beta_{n,m,k}^{*(1)},...,\beta_{n,m,k}^{*(R_{n,m}^*)})^T$.
\item \textbf{Restriction on the GLM densities:} Assume $F(\theta)=1+\theta\max\limits_{m=1:M}\{\sup_{|w|\leq \theta}|a_m'(w)|\sup_{|w|\leq \theta}|b_m'(w)/a_{m}'(w)|\}$ grows at most linearly with $\theta$. 
\item \textbf{Growth of node sparsity and hyper-parameters:} Following established literature, the hyper-parameter $b_{\eta}$ is assumed to be a function of $n$. The number of graph nodes $K_n$, the number of truly influential graph nodes $s_n$ and the hyper-parameter $b_{\eta}$ together satisfy
$b_{\eta}s_n/K_n\succ n$. For algebraic simplicity, $b_{\eta}$ is considered to be an integer without loss of generality.
\item \textbf{Bounded covariate:} The random covariate is bounded, i.e., there exists $a_0>0$ s.t. $|x|\leq a_0$.
\end{enumerate}
\textbf{Remark:} Assumption (D) is a general condition on ``model sparsity'' stating that most of the node-specific latent variables are small in magnitude. This is trivially satisfied when the true number of influential nodes is fixed and finite. Assumption (E) restricts the form of the GLM densities and is satisfied for a number of popular densities, e.g., when $f_{m,\bk}$ is a normal density with a known error variance and when $f_{m,\bk}$ corresponds to the density of a binary random variable $y_{n,m,\bk}$ with a logit or a probit link.
\begin{theorem}\label{theorem1}
Suppose Assumptions (A)-(G) hold. Denote the integrated Hellinger distance between the fitted and true density by $h(f,f^*)=\sqrt{\int\int(\sqrt{f(\by|x,\bgamma_n)}-\sqrt{f(\by|x,\bgamma_n^*)})^2\nu_{x}(dx)\nu_{\by}(d\by)}$, where $\nu_{\by}(d\by)$ and $\nu_x(dx)$ are dominating measures for $\by$ and $x$, with $\nu_x(dx)$ being the empirical measure for the data. Then 
$\Pi(f:h(f,f^*)\geq \epsilon|\by_n,x_1,...,x_n)\rightarrow 0$ as $n\rightarrow\infty$.
\end{theorem}
Theorem~\ref{theorem1} shows convergence of the predictive density from the proposed model to the true model. Next, we state a result to show that the posterior probability of the number of nodes identified as influential being larger than a constant multiple of the true number of influential nodes asymptotically vanishes. For simplicity, we consider $a_{m}'(w)=\sigma_m^{*2}$ (with $\sigma_m^{*2}$ known) and $b_m'(w)/a_m'(w)=-w$ for the result, which corresponds to $f_{m,\bk}$ being a normal distribution with known error variance $\sigma_m^{*2}$. Without loss of generality, we set $\sigma_m^{*2}=1$. Let $\tilde{\zeta}=\left\{k\in\{1,...,K_n\}:\tilde{\bbeta}_{n,k}\neq\bzero\right\}$ denote the node-indices corresponding to nonzero node-specific latent variables, such that $|\tilde{\zeta}|$ denotes the number of influential nodes. Assume $\tilde{\bbeta}_{n,m,k}^*=(\beta_{n,m,k}^{(1)},...,\beta_{n,m,k}^{(R_{n,m}^*)})^T$ encodes the true relationship between the key predictor and $k$th node in the $m$th graph response, and, let $\tilde{\bbeta}_{n,k}^*=(\tilde{\bbeta}_{n,1,k}^{*T},...,\tilde{\bbeta}_{n,M,k}^{*T})^T$ be the overall effect of the $k$th node on the $M$ graph views. 
%Further, denote 
%$\bgamma_{n,m,\zeta^*}^*=(\gamma_{n,m,\bk}^*:k_1,k_2\in\tilde{\zeta}^*)^T$ is the vector of nonzero true coefficients and $\bgamma_{n,m,\zeta^*}=(\gamma_{n,m,\bk}:k_1,k_2\in\tilde{\zeta}^*)^T$
%is the vector of fitted coefficient corresponding to nonzero true coefficients. 
\begin{theorem}\label{theorem2}
Suppose $\mathcal{C}_n=\{|\tilde{\zeta}|>C_0s_n\}$ denote the set  such that the number of influential nodes is greater than a constant multiple of the true number of influential nodes, where $C_0>0$ is a constant free of $n$. Then, for sufficiently large $C_0$, $E_{\bgamma_n^*}\Pi(\mathcal{C}_n|\by_n,x_1,...,x_n)\rightarrow 0$, as $n\rightarrow\infty$, under Assumptions (A), (B), (C), (D), (E), (F) and (G).
\end{theorem}
The proofs of both Theorems~\ref{theorem1} and \ref{theorem2} are presented in Appendix A.

\section{Posterior Computation}\label{sec:post_comp}
Since the full conditional distributions of all parameters follow standard families, posterior computation proceeds via Gibbs sampling. The supplementary file provides detailed information on the full conditional distributions. Our code implementation is in \texttt{R}, and it runs on a cluster computing environment with three interactive analysis servers. Each server is equipped with 56 cores and features the Dell PE R820: 4x Intel Xeon Sandy Bridge E5-4640 processor, 16GB RAM, and 1TB SATA hard drive.

In Section~\ref{sec:simulation}, we conduct various simulation experiments, each with $M=2$ graph views, and including scenarios with each view featuring continuous edges between nodes (i.e., continuous cell entries of the symmetric matrix corresponding to each graph view). The MCMC sampler iterates for $5000$ steps, discarding the initial $1000$ steps as burn-in and drawing posterior inference using the post-burn-in samples with a thinning factor of 2. In the posterior computation, each iteration does not require inverting a matrix larger than $RM \times RM$, leading to rapid computation. For instance, with an unoptimized code, the time to compute $5000$ MCMC iterations for cases with $K=40$ and $K=80$ nodes (both with $n=150$) was approximately $3.31$ hours and $9.83$ hours, respectively. Further optimization may be possible by parallelizing the draw over node-specific latent variables.

\section{Simulation Experiments}\label{sec:simulation}
This section assesses the inferential performance of the proposed joint model for multiview graphs, denoted as the joint learning (JL) framework, in comparison to competing methods using synthetic graphs generated across different simulation scenarios. For each simulation scenario, we evaluate the capability of each competitor to accurately identify nodes associated with a key predictor and to precisely estimate coefficients corresponding to each predictor, while providing uncertainty quantification (UQ).

\subsection{Simulation Settings, Competitors and Metrics of Comparison}
All simulation settings assume that the number of views is $M = 2$ and that the $m$th graph view is simulated using
\begin{align}\label{eq:sim_model}
E[y_{i,m,\bk}] =G_m^{-1}(\mu_m^{(0)}+x_{i}\gamma_{m,\bk}^{(0)}+ \tilde{x}_{i}\alpha^{(0)}_{m}),\:i = 1,...,n;\:\:m \in \{1,2\},\: \bk\in\mathcal{K}.
\end{align}
Our simulation setting assumes both 
$G_1(\cdot)$ and $G_2(\cdot)$ as identity links,  i.e., edge weights from both graph views are continuous. So the data generating model becomes
$y_{i,m,\bk} =\mu_m^{(0)}+x_{i}\gamma_{m,\bk}^{(0)}+ \tilde{x}_{i}\alpha^{(0)}_{m}+\epsilon_{i,m,\bk}$,
where $\epsilon_{i,m,\bk} \sim N(0, \sigma_m^2)$.
%We consider two simulation settings using different link functions and considering edge weights to be binary or continuous.\\
%\textbf{Simulation 1.} This simulation setting assumes both 
%$G_1(\cdot)$ and $G_2(\cdot)$ as identity links,  i.e., edge weights from both graph views are continuous. So the data generating model becomes
%$y_{i,m,\bk} =\mu_m^{(0)}+x_{i}\gamma_{m,\bk}^{(0)}+ \tilde{x}_{i}\alpha^{(0)}_{m}+\epsilon_{i,m,\bk}$,
%where $\epsilon_{i,m,\bk} \sim N(0, \sigma_m^2)$.\\ 
%\textbf{Simulation 2.} This simulation setting assumes 
%$G_1(\cdot)$ as the identity link and $G_2(\cdot)$ as the logit link, i.e., the edge weights from the first and second graph views are continuous and binary, respectively. 
Simulation settings assume $P=1$ and $\tilde{P} = 1$ in equation (\ref{eq:joint_network_model}), i.e., one key and one auxiliary predictor. We draw each element of the \emph{key predictor} vector $\bx = (x_{1}, x_{2},...,x_{n})^T$ and the \emph{auxiliary predictor} vector $\tilde{\bx} = (\tilde{x}_{1}, \tilde{x}_{2},...,\tilde{x}_{n})^T$ from $N(0,1)$. 

Let $\eta^{(0)}$ represent the probability of a node being influentially related to the key predictor, referred to as the \emph{node density} parameter. We simulate indicator variables that represent the activation status of the $K$ nodes, denoted $\xi_{1}^{(0)},...,\xi_{K}^{(0)}\stackrel{i.i.d.}{\sim} Ber(\eta^{(0)})$. When the $k$th node $\mathcal{N}_k$ is uninfluential (i.e., $\xi_{k}^{(0)}=0$), we set the $k$th node-specific latent variables, each of dimension $R^{(0)}$, corresponding to the two graphs as $\tilde{\bbeta}_{k}^{(0)}=(\tilde{\bbeta}_{1,k}^{(0)},\tilde{\bbeta}_{2,k}^{(0)})^T=\bzero$. On the other hand, when $\xi_{k}^{(0)}=1$, $\tilde{\bbeta}_{k}^{(0)}$ is generated from $N_{2R^{(0)}}(\bmu_{\bbeta}^{(0)},\bSigma_{\bbeta}^{(0)})$. Each entry of $\bmu_{\bbeta}^{(0)}$ is generated from N(0,1), each diagonal entry of 
$\bSigma_{\bbeta}^{(0)}$ is set to 1, and each off-diagonal entry is fixed at $0.5$. The covariance structure $\bSigma_{\bbeta}^{(0)}$ introduces predictor-dependent associations between the two graphs through node-specific latent vectors. The $\bk = (k_1, k_2)$th entry, $\gamma^{(0)}_{m, \bk}$ for the true coefficient matrix $\bGamma_m^{(0)} \in \mathbb{R}^{K \times K}$ for the key predictor corresponding to the $m$th graph is given by $\frac{({\bbeta_{m,k_1}^{(0)}})^T\bbeta_{m,k_2}^{(0)}}{2}$. Thus, $\gamma^{(0)}_{m, \bk}\neq 0$, if and only if both $\mathcal{N}_{k_1}$ and $\mathcal{N}_{k_2}$ are related to the key predictor (i.e., $\xi_{k_1}^{(0)}=\xi_{k_2}^{(0)}=1$).

We explore various scenarios in simulations %both \textbf{Simulation 1} and \textbf{Simulation 2} 
to provide a comprehensive assessment of the model's performance. In each scenario, the graph coefficients are generated by modifying the node density parameter $\eta^{(0)}$ and the dimension $R^{(0)}$ of the true node-specific latent variable. The fitted dimension of the latent variable is $R$, and we present different model fitting scenarios with $R > R^{(0)}$. For all simulations, the number of observations ($n$) and the number of graph nodes ($K$) are fixed at $150$ and $40$, respectively. Consequently, the number of upper triangular cell entries of each view of the graph matrix is given by $\mathcal{N}(\mathcal{K}) = K(K-1)/2 = 780$. The true coefficients for the auxiliary predictors are set as $\alpha^{(0)}_1 = 0.4$ and $\alpha^{(0)}_2 = -0.1$ for the first and second graph view, respectively. The true intercept terms for the two graphs, denoted by $\mu_1^{(0)}$ and $\mu_2^{(0)}$, are $0.2$ and $0.8$, respectively. %In \textbf{Simulation 1}, 
The true error variances are $\sigma_1^2 = 1$ and $\sigma_2^2 = 0.5$. %while \textbf{Simulation 2} sets the true error variance $\sigma_1^2 = 1$ for the graph with continuous edge weights.
Table \ref{Tab1} provides the specifications for the six scenarios in %both \textbf{Simulation 1} and \textbf{Simulation 2} 
by varying  $(\eta^{(0)},R^{(0)},R)$.

\noindent\textbf{Metrics of comparison.} For Scenarios 1--6, %in both Simulation 1 and Simulation 2, 
we assess parameter estimation accuracy using the mean squared error (MSE) metric. Since both the estimated graph coefficient $\bGamma_m$ and the true coefficient $\bGamma^{(0)}_m$ are symmetric matrices, the MSE for the $m$th graph coefficient, denoted $MSE_m$, is computed as
$MSE_m = \frac{2}{K(K-1)} \sum_{k_1 < k_2} \left( \hat{\gamma}_{m,k_1,k_2} - \gamma^{(0)}_{m,k_1,k_2} \right)^2,$
where $\hat{\gamma}_{m,k_1,k_2}$ is the point estimate of $\gamma_{m,k_1,k_2}$. In Bayesian models such as ours, $\hat{\gamma}_{m,k_1,k_2}$ represents the posterior mean. The effectiveness of uncertainty quantification (UQ) in parameter estimation is evaluated by the coverage probability and average length of the $95\%$ credible intervals for the edge coefficients, denoted as $Coverage(m)$ and $Length(m)$, respectively, for each graph view $m \in \{1,2\}$. Additionally, we report the area under the ROC curve (AUC) to assess the identification of influential graph nodes associated with the predictor. Higher AUC values reflect greater accuracy and reduced uncertainty in detecting influential nodes.

\noindent\textbf{Competitors.} As a competing approach to our proposed predictor-dependent joint learning (JL) framework, we consider treating the multiview graph as a $K\times K\times 2$ dimensional tensor and applying a predictor-dependent tensor learning (TL) approach \citep{guhaniyogi2021bayesian}. This method allows for point estimation and uncertainty quantification (UQ) of the tensor coefficient associated with the key predictor; however, it does not facilitate inference on influential graph nodes. In addition, we apply predictor-dependent modeling to each graph separately, as described in \cite{guha2021bayesiantensor}. We refer to this as the \emph{independent learning} (IL) approach. Since IL models each graph independently, it can identify different sets of nodes associated with the key predictor in each graph. Therefore, our comparison between IL and JL focuses on coefficient inference, in order to evaluate the benefits of joint versus independent coefficient estimation. %Both TL and IL are included as competing methods in Simulation 1. However, the TL approach, as in \cite{guhaniyogi2021bayesian}, is implemented assuming all tensor entries are continuous; thus, it is not applicable in Simulation 2, where the second graph contains binary edge weights. 

\subsection{Simulation Results} 
\begin{figure}[!ht]
  \begin{center}
  \includegraphics[width=13cm,height=12cm]{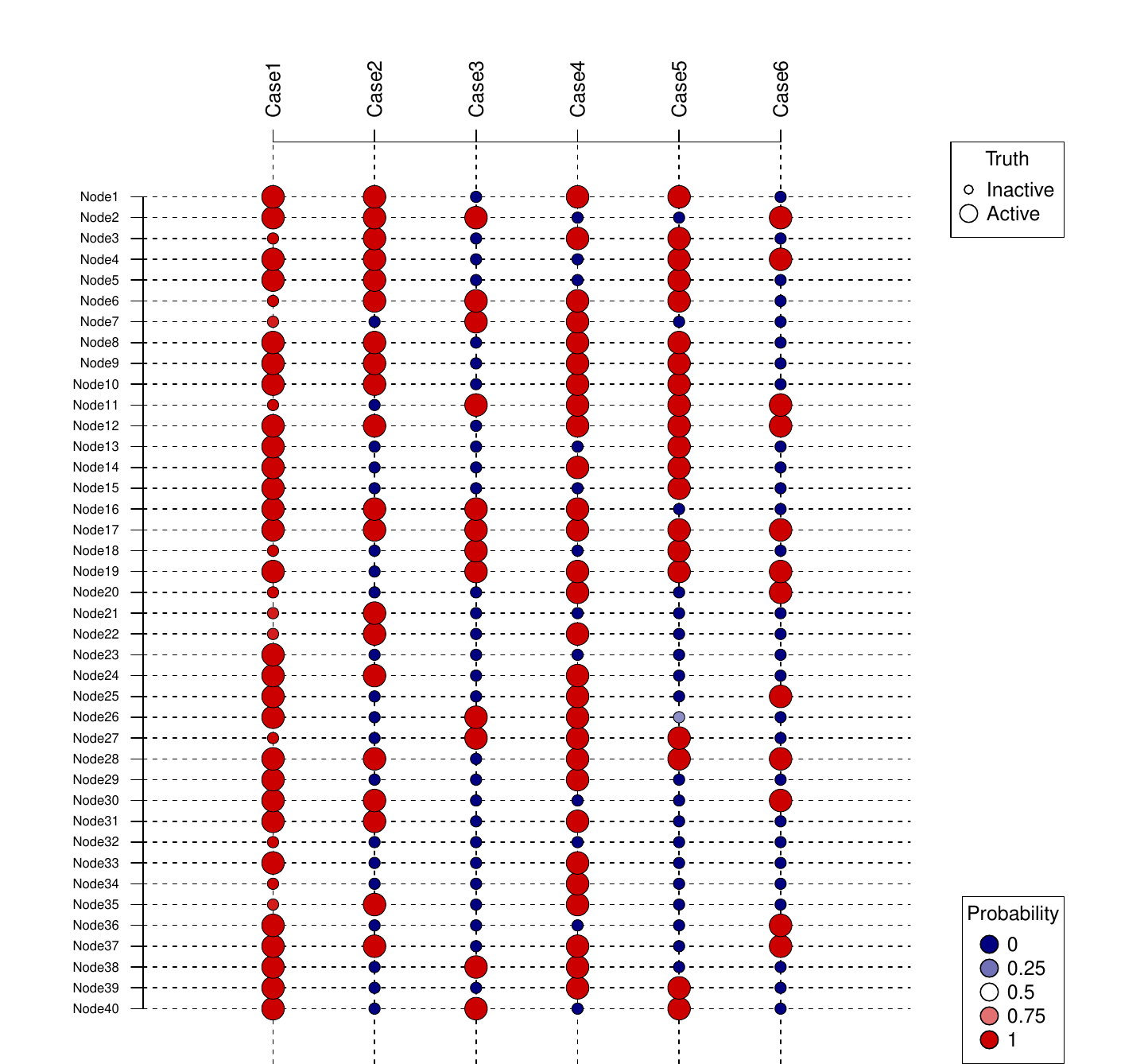}
 \end{center}
 \caption {Node activity level as detected by the model. The nodes are indicated along the rows and simulation cases are indicated along the columns. Within the figure, a bigger circle denotes a truly active node, while a smaller circle denotes a truly inactive node. The color spectrum moves from blue to red as the posterior probability of node activity  increases. The figure indicates excellent performance in terms of node detection. } 
\label{Sim1-ROIs}
\end{figure}

Figure~\ref{Sim1-ROIs} displays the node activity levels detected by the model for different simulation settings. Figure~\ref{Sim1-ROIs} suggests that the model accurately estimates the posterior probability of a node being influential, i.e., $P(\xi_k=1|\mbox{Data})$, close to 1 or 0 when the node is truly influential or truly un-influential, respectively. 
Shifting our focus to the network coefficients, we observe excellent point estimation for both network coefficients $\bGamma_m^{(0)}$, $m=1,2$, under all scenarios for the predictor-dependent joint learning (JL) approach. When keeping the rank of the true coefficient $R^{(0)}$ fixed, a decrease in node density generally leads to a lower mean squared error (MSE). As we increase $R^{(0)}$, fixing the true sparsity, the point estimation marginally deteriorates. There is no notable effect on the inference due to the user-specific choice of the fitted rank $R$ when $R$ is kept within a moderate range. This lack of sensitivity is attributed to the shrinkage imposed by the model on the utilization of higher ranks. 
\begin{table}[!h]
\centering
\begin{tabular}
[c]{c|cccccc|cccccc}
\cline{1-13}
MSE $\times 10^2$ & \multicolumn{6}{ c| }{\textbf{Simulation 1} ($m=1$)} & \multicolumn{6}{ c }{\textbf{Simulation 1} ($m=2$)}\\ 
 &  &  &  &  &  &  &  &  &  &  &  &  \\
\hline
% &  &  &  &  &  &  &  &  &  &  &  &  \\
$\eta^{(0)}$ & 0.7 & 0.5 & 0.3 & 0.7 & 0.5 & 0.3 & 0.7 & 0.5 & 0.3 & 0.7 & 0.5 & 0.3\\
$R^{(0)}$ & 4 & 4 & 4 & 3 & 3 & 3 & 4 & 4 & 4 & 3 & 3 & 3\\
$R$ & 5 & 8 & 8 & 5 & 8 & 8 & 5 & 8 & 8 & 5 & 8 & 8\\
Scenarios & 1 & 2 & 3 & 4 & 5 & 6 & 1 & 2 & 3 & 4 & 5 & 6\\
 &  &  &  &  &  &  &  &  &  &  &  &  \\
\hline
% &  &  &  &  &  &  &  &  &  &  &  &  \\
JL & \textbf{0.08} & \textbf{0.07} & \textbf{0.03} & \textbf{0.08} & \textbf{0.05} & \textbf{0.02} & \textbf{0.17} & \textbf{0.18} & \textbf{0.10} & \textbf{0.14} & \textbf{0.12} & \textbf{0.09}\\
IL & 0.14 & 0.12 & 0.05 & 0.13 & 0.13 & 0.06 & 0.24 & 0.26 & 0.14 & 0.22 & 0.18 & 0.15 \\
TL & 0.59 & 0.64 & 0.47 & 0.65 & 0.57 & 0.52 & 0.84 & 0.76 & 0.73 & 0.69 & 0.72 & 0.63\\
 &  &  &  &  &  &  &  &  &  &  &  &  \\
\hline
\end{tabular}
\caption{\label{Tab1} Table presents mean squared errors (MSE)$\times 10^2$ for Scenarios 1-6. Here, $m=1$ and $m=2$ correspond to the first and second response networks, respectively. The parameter $\eta^{(0)}$ refers to the probability of a node being active (\emph{node density}). Different cases present various combinations of network node density $(\eta^{(0)})$, true ($R^{(0)}$) and fitted  ($R$) dimensions of the node-specific latent variables.}
%\end{center}
\end{table}
In all simulations, JL demonstrates superior performance over independent learning (IL), with the performance gap widening with increasing node density. This is justifiable since IL does not account for the correlation between the two networks through the joint modeling of node-specific latent variables. The impact of ignoring this correlation structure is most apparent when more nodes are set active in generating the two networks. Tensor learning (TL) shows significantly inferior performance compared to its competitors, perhaps due to ignoring the symmetry in each of the two slices, represented by the two networks, of the tensor response.

\section{Functional Connectivity under Diverse Cognitive Control Domains}\label{sec:real_data_analysis}

In this section, we analyze functional connectivity data from functional magnetic resonance imaging (fMRI) of 144 healthy adults (ages 20–86) in the greater Toronto area, collected during performance of cognitive control tasks \citep{rieck2021reconfiguration}. During scanning on a Siemens 3T MRI, participants performed a go/no-go task that assessed 2 tasks, namely \textbf{inhibition} and \textbf{initiation}: responding to the letter ``X'' (go) and refraining from responding to other letters (no-go). The task included blocks favoring go trials (inhibition: 120 go, 40 no-go) and blocks favoring no-go trials (initiation: 20 go, 60 no-go), with block order randomized.
Letter stimuli were displayed for 400 ms with an average inter-stimulus interval of 1200 ms, varying between 900 and 1500 ms. The entire go/no-go task lasted 6 minutes and 24 seconds. Blood-oxygen-level dependent (BOLD) fMRI data were collected using a 12-channel head coil with an echo-planar imaging sequence. %, acquiring 40 axial slices parallel to the anterior-posterior commissure. 
For the go/no-go task, 216 volumes were collected. High-resolution anatomical scans for warping the BOLD images to MNI space were obtained with a T1-weighted MP-RAGE sequence with 160 axial slices. Functional data were preprocessed using a mix of AFNI functions, Octave, and MATLAB scripts with the Optimizing of Preprocessing Pipelines for NeuroImaging (OPPNI) software package  \citep{churchill2017optimizing}. The preprocessing steps included: (1) rigid-body alignment to correct for movement; (2) removal and interpolation of outlier volumes; (3) correction for physiological noise (cardiac and respiratory); (4) slice timing correction; (5) spatial smoothing with a 6 mm smoothing kernel; (6) temporal de-trending; (7) regression of six motion parameter estimates (X, Y, Z translation and rotation) on the time-series; (8) regression of signal in non-interest tissues (white matter, vessels, cerebrospinal fluid) on the time-series; and (9) warping to MNI space and resampling to $4 mm^3$ isotropic voxels.

\begin{figure}
    \centering
    \includegraphics[width=0.95\linewidth]{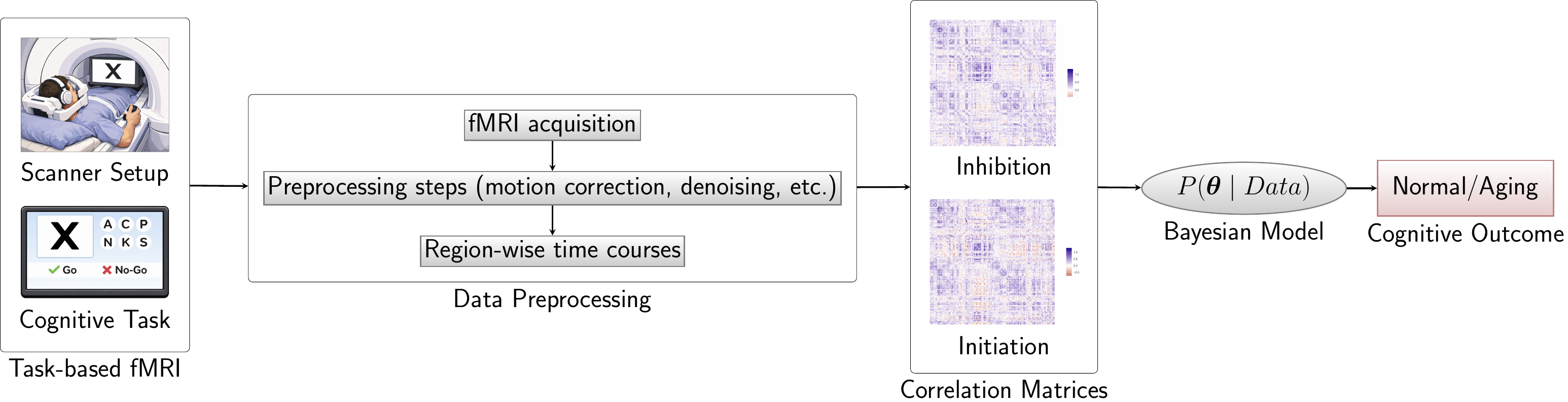}
    \caption{Visualization of the steps for fMRI multiview graph data acquisition and modeling.}
    \label{real_data_flowchart}
\end{figure}

Subsequently, for each participant, a multiview graph comprising two views, corresponding to (1) \textbf{inhibition}, and (2) \textbf{initiation} task conditions, was constructed. Each graph view consisted of 200 nodes, representing brain regions of interest (ROIs) defined by the Schaefer 200-parcel 17-network atlas \citep{schaefer2018local}. The connection strength between each pair of regions was calculated as the Z-transformed correlation coefficient of their BOLD time series, resulting in a $200\times 200$ functional connectivity matrix for each task condition. These functional connectivity matrices were computed to investigate the neurobiology of aging. Figure~\ref{real_data_flowchart} represents the flowchart for the multiview graph data acquisition and modeling steps. The key predictor is the Mini-Mental State Examination (MMSE) score \cite{folstein1975, mega1996spectrum}. MMSE is a widely used measure for detection of cognitive impairment by assessing functions such as orientation, memory, attention, and language. A lower score is indicative of greater impairment. Age and sex serve as auxiliary predictors.

\subsection{Prediction Results}
 Table~\ref{real_data_prediction} summarizes the point predictions and predictive uncertainty, averaged across the two fMRI graph views, for the competing methods. Consistent with the findings from the simulation studies, tensor learning demonstrates the poorest performance, which may be attributable to its inability to account for the inherent symmetry in each graph view. Joint learning continues to outperform independent learning, although the margin of improvement is less pronounced than what was observed in the simulations. Similar to simulations, tensor learning performs inferior to both joint learning and independent learning.
\begin{table}[H]
    \centering
    \begin{tabular}{c|c c c}
        Model & MSPE & Interval Coverage & Interval Length \\
         \hline
         JL & 0.4796 & 0.9582 & 1.5545 \\
         \hline
        IL & 0.5304 & 0.9462 & 1.4860 \\ 
         \hline
        TL & 0.5833 & 0.9023 & 0.9612 \\
    \end{tabular}
    \caption{Mean Squared Prediction Error (MSPE) and 95\% Prediction Interval Coverages and Lengths for the Joint Learning (JL), Independent Learning (IL), and Tensor Learning (TL) approaches. Results are averaged across the two fMRI graph views.}
    \label{real_data_prediction}
\end{table}

%\textcolor{blue}{Mention the selected nodes and write a small scientific justification. It is Bayesian Analysis so we do not need a lot of details.}

Our method identified 88 significant brain regions of interest (ROIs) from the fMRI multiview graphs associated with MMSE performance, with 42 located in the left hemisphere and 46 in the right. Regions of interest were selected for inclusion if their posterior inclusion probability $P(\xi_k=1|\mbox{Data})$ exceeded 0.5, a threshold chosen according to the median probability rule \citep{barbieri2004optimal}. Table~\ref{Selected_ROI_Counts_Total_Proportions_7_network} reports the counts and proportions of selected ROIs within each functional network, and the complete list of selected ROIs is provided in Table 1 of Appendix C.

At the network level, selected ROIs were concentrated in higher-order association networks, with the dorsal attention, control, and salience/ventral attention networks exhibiting higher within-network selection proportions (Table~\ref{Selected_ROI_Counts_Total_Proportions_7_network}). The limbic network also showed a relatively high proportion of selected ROIs. In contrast, the default mode network (DMN) exhibited a lower selection proportion relative to other association networks, and primary sensory and motor systems, including the visual and somatomotor networks, also showed lower selection proportions. This distribution aligns with prior neuroimaging findings linking executive and attentional association networks, as well as the limbic system, to age-related differences in cognitive control \citep{AndrewsHanna2007, Chan2014, Dolcos_2012}.

The prominence of dorsal attention, control, and salience/ventral attention network ROIs aligns with their established roles in inhibitory control and attentional regulation during go/no-go paradigms, and with evidence that age-related declines in executive function are associated with altered recruitment of these systems \citep{Corbetta_2002, Rubia2001, Verbruggen2008,Vossel_2014}. Furthermore, the substantial representation of the limbic network supports its involvement in emotion regulation, motivation, and memory processes that interact with executive control demands \citep{Pessoa_2008, Samanez-Larkin_2015}.

Default mode network ROIs were selected at a lower rate relative to other higher-order networks, consistent with their reduced engagement during externally directed cognitive control, despite evidence linking age-related alterations in DMN activity and suppression to cognitive decline \citep{Anticevic_2012, Grady2010, Spreng2016}. Visual and somatomotor ROIs, while directly engaged by task stimuli and motor responses, also showed lower selection proportions. This pattern is consistent with prior work suggesting that age-related cognitive differences are more strongly associated with variability in large-scale association and limbic networks than with variability in primary sensory and motor systems \citep{Cabeza_2018, Park_2004}.

Overall, these results demonstrate that the joint learning approach preferentially identifies connectivity patterns within networks supporting executive, attentional, and affective–mnemonic processes, reinforcing the biological plausibility and interpretability of the framework for understanding MMSE-related cognitive performance.

\begin{table}[H]
    \centering
    \begin{tabular}{c|c|c|c}
        Network & Selected ROIs & Total ROIs & Proportion Selected \\
        \hline
        Dorsal Attention Network & 12 & 22 &  0.55 \\
        Control Network & 20  & 37 & 0.54 \\
        Salience/Ventral Attention Network & 14 &  26 & 0.54  \\
        Limbic Network & 7 & 14 &  0.50 \\
        Somatomotor Network & 12 & 34 &   0.35  \\
        Default Mode Network & 15 & 43 &  0.35 \\ 
         Visual Network & 8 & 24 &  0.33  \\
    \end{tabular}
    \caption{Distribution of selected ROIs across functional networks in the Schaefer 200-parcel atlas. Networks are ordered by decreasing within-network selection proportion. The proportion selected reflects the fraction of ROIs within each network identified by the joint learning procedure as associated with MMSE performance.}
    \label{Selected_ROI_Counts_Total_Proportions_7_network}
\end{table}

\section{Conclusion and Future Work}\label{sec:conclusions}
In this work, we propose a novel hierarchical Bayesian modeling framework for predictor-dependent joint learning of multiview graphs on a common set of nodes. Our approach quantifies the relationship between edge structures and predictors, and facilitates rigorous model-based inference for identifying nodes significantly associated with a given predictor. The hierarchical Bayesian formulation naturally provides uncertainty quantification for all inferential statements. We established theoretical guarantees, including Bayesian asymptotic results that demonstrate convergence of the fitted predictive density of the predictor-dependent multiview graph model to the true density under mild regularity conditions. In both simulated and functional connectivity datasets, our framework outperformed competing methods such as independent predictor-dependent learning of graph views and predictor-dependent tensor learning. 

While our work is motivated by functional connectivity (FC) datasets where spatial locations of nodes are unavailable, some emerging applications involve graphs with spatially localized nodes. A natural direction for future research is to formulate a framework to incorporate spatial information, enabling spatial smoothing or borrowing of strength across neighboring nodes. Another important avenue is to further advance the modeling framework by developing semi-parametric approaches that can capture nonlinear relationships between multiview graphs and predictors, while preserving the ability to conduct node-level inference. These extensions have strong potential to increase the flexibility and applicability of predictor-dependent joint learning of multiview graphs in complex real-world datasets.

\bibliographystyle{natbib}
\bibliography{references}

\end{document}

\maketitle

\begin{abstract}
This supplementary material contains two appendices. Appendix A contains the proofs of theorems 3.1 and 3.2. Appendix B details the full conditional distributions used for Gibbs sampling. Appendix C 
offers the list of regions of interest identified as being associated with the mini mental state examination score.
\end{abstract}

\newpage

\section*{Appendix A}
To prove the theorem, we first state and prove a series of lemmas. The results from these lemmas will be used to prove Theorem 1 and 2. 
\begin{lemma}\label{lemma_app1}
Assume for $r=1,...,R_{n}$, and $\omega>1$, $\lambda_m^{(r)}$ takes values $0,1,-1$ with probabilities $\pi_{m,1}^{(r)},\pi_{m,2}^{(r)},\pi_{m,3}^{(r)}$, respectively. Let 
$(\pi_{m,1}^{(r)},\pi_{m,2}^{(r)},\pi_{m,3}^{(r)})$ follow a Dirichlet($r^\omega,1,1$). Then
%$r=1,...,R_n$, and $\eta>0$,
\begin{align*}
P(\mathcal{F})=P(\lambda_m^{(1)}=1,...,\lambda_m^{(R_{n})}=1)
\geq \frac{1}{(2+R_{n}^{\omega})^{R_{n}}}.
\end{align*}
\end{lemma}
\begin{proof}
$P(\lambda_m^{(r)}=1)=E(\pi_{m,2}^{(r)})=\frac{1}{2+r^{\omega}}$ for $r=1,...,R_{n}$. Then,
\begin{align*}
 P(\lambda_m^{(1)}=1,...,\lambda_m^{(R_n)}=1)
=\prod_{r=1}^{R_{n}}\frac{1}{(2+r^{\omega})}
\geq \frac{1}{(2+R_{n}^{\omega})^{R_n}}.
%\left\{\frac{R_{n,m}^{*\omega}}{(2+R_{n,m}^{*\omega})}\right\}^{R_{n}-R_{n,m}^*}
%=\frac{2^{R_{n,m}^*}R_{n,m}^{*\omega(R_{n}-R_{n,m}^*)}}{(2+R_{n,m}^{*\omega})^{R_{n}}}.
\end{align*}
The inequality follows from the fact that %$r^{\omega}/(2+r^{\omega})$ is a monotone increasing function of $r$ and
$1/(1+r^{\omega})$ is a monotone decreasing function of $r$.
\end{proof}

\begin{lemma}\label{lem1}
Let Assumptions (B) and (C) hold, and, define $\tilde{\bbeta}_{n,k}=(\tilde{\bbeta}_{n,1,k}^T,...,\tilde{\bbeta}_{n,M,k}^T)^T$, where $\tilde{\bbeta}_{n,m,k}=(\beta_{n,m,k}^{(1)},...,\beta_{n,m,k}^{(R_{n})})^T$. For $R_{n}>R_{n,m}^*$, define $\tilde{\bbeta}_{n,m,k}^*=(\beta_{n,m,k}^{*(1)},...,\beta_{n,m,k}^{*(R_{n})})^T$, with $\beta_{n,m,k}^{*(r)}=0$, for $r=R_{n,m}^*+1,...,R_{n}$. Then for any $\Delta_n>0$,
\begin{align}\label{eq_bound}
& -\log\Pi(||\tilde{\bbeta}_{n,k}-\tilde{\bbeta}_{n,k}^*||_2\leq \Delta_n;\:\:k=1,...,K_n)\nonumber\\
&\leq \sum_{k=1}^{K_n}\sum_{m=1}^M\sum_{r=1}^{R_{n,m}^*}\frac{|\beta_{n,m,k}^{*(r)}|^2}{2}+\frac{R_nK_nM}{2}\log(2\pi)+\log\left(R_nK_nM+1\right)+\frac{R_nK_nM}{2}\log\left(R_nM\right)\nonumber\\
&\qquad\qquad\qquad\qquad\qquad+R_nK_nM+1-R_nK_nM\log\left(\Delta_n\right)+\frac{K_n\Delta_n^2}{4}.
\end{align}
\end{lemma}
\begin{proof}
Note that
\begin{align*}
&\Pi(||\tilde{\bbeta}_{n,k}-\tilde{\bbeta}_{n,k}^*||_2\leq \Delta_n;\:\:k=1,...,K_n)
=\prod_{k=1}^{K_n}E\left[\Pi(||\tilde{\bbeta}_{n,k}-\tilde{\bbeta}_{n,k}^*||_2\leq \Delta_n|\eta)\right]\nonumber\\
&\geq \prod_{k=1}^{K_n}E[\exp(-||\tilde{\bbeta}_{n,k}^*||_2^2/2)\Pi(||\tilde{\bbeta}_{n,k}||_2\leq \Delta_n/2|\eta)]\nonumber\\
&=\exp(-\sum_{k=1}^{K_n}||\tilde{\bbeta}_{n,k}^*||_2^2/2)\prod_{k=1}^{K_n}E[\Pi(||\tilde{\bbeta}_{n,k}||_2\leq \Delta_n/2|\eta)],
\end{align*}
where the second inequality follows from the Anderson's Lemma \citep{guhaniyogi2017bayesian}. Now we proceed to develop a lower bound on $\Pi(||\tilde{\bbeta}_{n,k}||_2\leq \Delta_n/2|\eta)$. Indeed,
\begin{align*}
&\Pi(||\tilde{\bbeta}_{n,k}||_2\leq \Delta_n/2|\eta)
\geq \prod_{m=1}^M\prod_{r=1}^{R_{n}}\Pi\left(|\beta_{n,m,k}^{(r)}|\leq \frac{\Delta_n}{2\sqrt{R_{n}M}}\given\eta\right)\\
&=\prod_{m=1}^M\prod_{r=1}^{R_{n}}\left[(1-\eta)+\frac{\eta}{\sqrt{2\pi}}\int_{-\Delta_n/2\sqrt{R_{n}M}}^{\Delta_n/2\sqrt{R_{n}M}}e^{-\frac{x^2}{2}}dx\right]
\geq \left[(1-\eta)+\frac{\Delta_n\eta}{\sqrt{2R_nM\pi}}\exp\left(-\frac{\Delta_n^2}{4R_{n}M}\right)\right]^{R_{n}M},
\end{align*}
where the last inequality follows from the fact that 
$\int_{-b}^{b} e^{-x^2/2}dx\geq e^{-b^2}2b$, for any $b>0$.
Hence,
\begin{align*}
&\prod_{k=1}^{K_n}E[\Pi(||\tilde{\bbeta}_{n,k}||_2\leq \Delta_n|\eta)]
\geq E\left[(1-\eta)+\frac{\Delta_n\eta}{\sqrt{2R_nM\pi}}\exp\left(-\frac{\Delta_n^2}{4R_nM}\right)\right]^{R_nK_nM}\\
&=E\left[\sum_{h=1}^{R_nK_nM}{R_nK_nM \choose h}(1-\eta)^h
\left(\frac{\Delta_n\eta}{\sqrt{2R_nM\pi}}\right)^{R_nK_nM-h}\exp\left(-\frac{(R_nK_nM-h)\Delta_n^2}{4R_nM}\right)\right]\\
&=\sum_{h=1}^{R_nK_nM}{R_nK_nM \choose h}Beta(R_nK_nM-h+1,h+b_{\eta}+1)
\left(\frac{\Delta_n}{\sqrt{2R_nM\pi}}\right)^{R_nK_nM-h}\\
&\qquad\qquad\exp\left(-\frac{(R_nK_nM-h)\Delta_n^2}{4R_nM}\right)\\
&=\sum_{h=1}^{R_nK_nM}\frac{R_nK_nM!}{h!(R_nK_nM-h)!} \frac{(R_nK_nM-h)!(h+b_{\eta})!}{(R_nK_nM+b_{\eta}+1)!}
\left(\frac{\Delta_n}{\sqrt{2R_nM\pi}}\right)^{R_nK_nM-h}\\
&\qquad\qquad\exp\left(-\frac{(R_nK_nM-h)\Delta_n^2}{4R_nM}\right)\\
&\geq \left(\frac{1}{\sqrt{2\pi}}\right)^{R_nK_nM}\left\{\frac{1}{R_nK_nM+1}\prod_{k=1}^{b_{\eta}}\left(1+\frac{R_nK_nM+1}{k}\right)^{-1}\right\}
\left(\frac{\Delta_n}{\sqrt{R_nM}}\right)^{R_nK_nM}\exp(-K_n\Delta_n^2/4)\\
%&\geq \left(\frac{1}{\sqrt{2\pi}}\right)^{R_nK_nM}\left\{\frac{R_nK_nM}{R_nK_nM+1}\prod_{k=1}^{b_{\eta}}\left(1+\frac{R_nK_nM+1}{k}\right)^{-1}\right\}
%\left(\frac{\Delta_n}{\sqrt{R_nM}}\right)^{R_nK_nM}\exp(-K_n\Delta_n^2/4)\\
&\geq \left(\frac{1}{\sqrt{2\pi}}\right)^{R_nK_nM}\left\{\frac{1}{R_nK_nM+1}\left(1+\frac{R_nK_nM+1}{b_{\eta}}\right)^{-b_{\eta}}\right\}
\left(\frac{\Delta_n}{\sqrt{R_nM}}\right)^{R_nK_nM}\exp(-K_n\Delta_n^2/4).
\end{align*}
Now we use the fact that 
$\left(1+\frac{R_nK_nM+1}{b_{\eta}}\right)^{-b_{\eta}}\geq \exp(-R_nK_nM-1)$.
Combining the earlier expressions with the last line, we obtain
\begin{align*}
\Pi(||\tilde{\bbeta}_{n,k}-\tilde{\bbeta}_{n,k}^*||_2\leq \Delta_n;\:\:k=1,...,K_n)&\geq \exp(-\sum_{k=1}^{K_n}||\tilde{\bbeta}_{n,k}^*||_2^2/2) \left(\frac{1}{\sqrt{2\pi}}\right)^{R_nK_nM}\frac{1}{R_nK_nM+1}\\
&\exp(-R_nK_nM-1)\left(\frac{\Delta_n}{\sqrt{R_nM}}\right)^{R_nK_nM}\exp(-K_n\Delta_n^2/4).
\end{align*}
Hence,
\begin{align*}
& -\log\Pi(||\tilde{\bbeta}_{n,k}-\tilde{\bbeta}_{n,k}^*||_2\leq \Delta_n;\:\:k=1,...,K_n)\\
&\leq \sum_{k=1}^{K_n}\sum_{m=1}^M\sum_{r=1}^{R_{n,m}^*}\frac{|\beta_{n,m,k}^{*(r)}|^2}{2}+\frac{R_nK_nM}{2}\log(2\pi)+\log\left(R_nK_nM+1\right)+\frac{R_nK_nM}{2}\log\left(R_nM\right)\\
&\qquad\qquad\qquad\qquad\qquad+R_nK_nM+1-R_nK_nM\log\left(\Delta_n\right)+\frac{K_n\Delta_n^2}{4}.
\end{align*}
\end{proof}
\begin{lemma}\label{lem2}
Let $\bgamma_n=(\bgamma_{n,1}^T,...,\bgamma_{n,M}^T)^T$ be the $MQ_n\times 1$ dimensional vector of coefficients and $\bgamma_{n}^*$ is the true value of $\bgamma_n$. Assume
$a_{m}'(w)=1$ and $b_m'(w)/a_m'(w)=-w$ which corresponds to $y_{n,m,\bk}$ following a normal distribution with unit variance.
that $P_{\bgamma_n^*}$ denotes the true probability distribution of $\by_n$. Then for every $\epsilon>0$,
\begin{align}\label{eq:lem1}
P_{\bgamma_n^*}\left(\left\{\by_n:\int \frac{f(\by_n|x,\bgamma_n)}{f(\by_n|x,\bgamma_n^*)} \pi(\bgamma_n)d\bgamma_n\leq \exp(- n\epsilon^2)\right\}\right)\rightarrow 0, \:\:\mbox{as}\:\:n\rightarrow\infty,
\end{align}
under Assumptions (A), (B), (C), (D) and (G).
\end{lemma}
\begin{proof}
For two densities $g_1,g_2$, denote $K(g_1,g_2)=\int g_1\log(g_1/g_2)$ and 
$V(g_1,g_2)=\int g_1((\log(g_1/g_2))-K(g_1,g_2))^2$.
Define 
\begin{align}
\mathcal{A}_{n}=\left\{E_x\sum_{i=1}^nK(f(\by_i|x_i,\bgamma_n^*),f(\by_i|x,\bgamma_n))\lesssim n\epsilon^2,\:\:E_x\sum_{i=1}^nV(f(\by_i|x_i,\bgamma_n^*),f(\by_i|x,\bgamma_n))\lesssim n\epsilon^2\right\}.
\end{align}
By Lemma~10 in \cite{ghosal2007convergence}, to show (\ref{eq:lem1}) it is enough to show that $\Pi(\mathcal{A}_{n})\gtrsim \exp(-C_2n\epsilon^2)$, for any constant $C_2>0$. With some algebra, we derive the following expressions,
\begin{align}\label{eq2}
%K(f(\by_i|x,\bgamma_n^*),f(\by_i|x,\bgamma_n)) &= \frac{1}{2}\left\{Q_n\sum_{m=1}^M\left(\frac{\sigma_m^2}{\sigma_{m}^{*2}}-1-\log\left(\frac{\sigma_m^2}{\sigma_{m}^{*2}}\right)\right) + x_i^2\sum_{m=1}^M\frac{||\bgamma_{n,m}-\bgamma_{n,m}^*||^2}{\sigma_m^2}\right\},\nonumber\\
%V(f(\by_i|x,\bgamma_n^*),f(\by_i|x,\bgamma_n)) &= \frac{1}{2}\sum_{m=1}^{M}\left(1-\frac{\sigma_m^2}{\sigma_{m}^{*2}}\right)^2 + x_i^2\sum_{m=1}^M\frac{\sigma_m^{*2}||\bgamma_{n,m}-\bgamma_{n,m}^*||^2}{\sigma_m^4}.
E_x\sum_{i=1}^n K(f(\by_i|x,\bgamma_n^*),f(\by_i|x_i,\bgamma_n)) &=  E_x\sum_{i=1}^n\frac{x_i^2}{2}||\bgamma_{n}-\bgamma_{n}^*||_2^2=nE_x[x^2/2]||\bgamma_{n}-\bgamma_{n}^*||_2^2,\nonumber\\
E_x\sum_{i=1}^n V(f(\by_i|x,\bgamma_n^*),f(\by_i|x_i,\bgamma_n)) &= E_x\sum_{i=1}^n x_i^2||\bgamma_{n}-\bgamma_{n}^*||_2^2
=nE_x[x^2]||\bgamma_{n}-\bgamma_{n}^*||_2^2,
\end{align}
given our simplifying assumption of $y_{n,m,\bk}$ following a normal distribution with unit variance. Following (\ref{eq2}) and Assumption (G), it in enough to show that 
$\Pi(||\bgamma_{n}-\bgamma_{n}^*||_2\leq \epsilon)\gtrsim\exp(-C_2n\epsilon^2)$. Let $\mathcal{D}_n=\{\lambda_1^{(1)}=1,...,\lambda_1^{(R_n)}=1,...,\lambda_M^{(1)}=1,...,\lambda_M^{(R_{n})}=1\}$. 
We will show $\Pi\left(\left\{||\bgamma_{n}-\bgamma_{n}^*||_2\leq \epsilon\right\}\cap\mathcal{D}_n\right)\gtrsim\exp(-C_2n\epsilon^2)$.
Under $\mathcal{D}_n$,
\begin{align}\label{eq3}
&||\bgamma_{n}-\bgamma_{n}^*||_2^2
=\sum_{\bk=(k_1,k_2)\in\mathcal{K}}\left\|(\gamma_{n,1,\bk},...,\gamma_{n,M,\bk})^T-(\gamma_{n,1,\bk}^*,...,\gamma_{n,M,\bk}^*)^T\right\|_2^2\nonumber\\
&=\sum_{\bk=(k_1,k_2)\in\mathcal{K}}\left\|\sum_{r=1}^{R_n}\left\{\left(\beta_{n,1,k_1}^{(r)}\beta_{n,1,k_2}^{(r)},...,\beta_{n,M,k_1}^{(r)}\beta_{n,M,k_2}^{(r)}\right)^T-\left(\beta_{n,1,k_1}^{*(r)}\beta_{n,1,k_2}^{*(r)},...,\beta_{n,M,k_1}^{*(r)}\beta_{n,M,k_2}^{*(r)}\right)^T\right\}\right\|_2^2,
\end{align}
where the last equality follows from Assumptions (B) and (C), and by acknowledging the fact that $(\beta_{n,1,k}^{*(r)},...,\beta_{n,M,k}^{*(r)})^T=\bzero$, for all $r=R_{n,m}^*+1,...,R_n$. Denote  $\tilde{\bbeta}_{n,k}=(\tilde{\bbeta}_{n,1,k},...,\tilde{\bbeta}_{n,M,k})^T$, where $\tilde{\bbeta}_{n,m,k}=(\beta_{n,m,k}^{(1)},...,\beta_{n,m,k}^{(R_{n,m}^*)})^T$. Following (\ref{eq3}),
\begin{align*}
&||\bgamma_{n}-\bgamma_{n}^*||_2^2
\leq R_n\sum_{\bk=(k_1,k_2)\in\mathcal{K}}\sum_{r=1}^{R_n}
||(\bbeta_{n,k_1}^{(r)}-\bbeta_{n,k_1}^{*(r)})\circ\bbeta_{n,k_2}^{(r)}+(\bbeta_{n,k_2}^{(r)}-\bbeta_{n,k_2}^{*(r)})\circ\bbeta_{n,k_2}^{*(r)}||_2^2\\
&\leq 2R_n\sum_{\bk=(k_1,k_2)\in\mathcal{K}}\sum_{r=1}^{R_n}
\left[||(\bbeta_{n,k_1}^{(r)}-\bbeta_{n,k_1}^{*(r)})\circ\bbeta_{n,k_2}^{(r)}||_2^2+||_2(\bbeta_{n,k_2}^{(r)}-\bbeta_{n,k_2}^{*(r)})\circ\bbeta_{n,k_2}^{*(r)}||_2^2\right]\\
&\leq 2R_n\sum_{\bk=(k_1,k_2)\in\mathcal{K}}\sum_{r=1}^{R_n}
\left[||(\bbeta_{n,k_1}^{(r)}-\bbeta_{n,k_1}^{*(r)})\circ\bbeta_{n,k_2}^{(r)}||_2^2+||(\bbeta_{n,k_2}^{(r)}-\bbeta_{n,k_2}^{*(r)})\circ\bbeta_{n,k_2}^{*(r)}||_2^2\right]\\
&\leq 2R_n\sum_{\bk=(k_1,k_2)\in\mathcal{K}}\left[||\tilde{\bbeta}_{n,k_1}-\tilde{\bbeta}_{n,k_1}^{*}||_2^2||\tilde{\bbeta}_{n,k_2}||_2^2+||\tilde{\bbeta}_{n,k_2}-\tilde{\bbeta}_{n,k_2}^{*}||_2^2||\tilde{\bbeta}_{n,k_2}^{*}||_2^2\right]\\
&\leq 2R_n\sum_{\bk=(k_1,k_2)\in\mathcal{K}}\left[||\tilde{\bbeta}_{n,k_1}-\tilde{\bbeta}_{n,k_1}^{*}||_2^2\left(2||\tilde{\bbeta}_{n,k_2}-\tilde{\bbeta}_{n,k_2}^*||_2^2+2||\tilde{\bbeta}_{n,k_2}^*||_2^2\right)+||\tilde{\bbeta}_{n,k_2}-\tilde{\bbeta}_{n,k_2}^{*}||_2^2||\tilde{\bbeta}_{n,k_2}^{*}||_2^2\right],
\end{align*}
where $\circ$ denotes the outer product between two vectors.
Define a set $\mathcal{H}_n$ given by
\begin{align}\label{eq:set}
\mathcal{H}_n=\left\{||\tilde{\bbeta}_{n,k}-\tilde{\bbeta}_{n,k}^*||_2^2\leq \upsilon_n;\:\:k=1,...,K_n,\:\mbox{where}\:\upsilon_n>1/n,\:\:4\upsilon_n^2R_nQ_n+6\upsilon_nR_nK_n\tilde{C}_{\beta}^2\leq \epsilon^2\right\}.
\end{align}
$\mathcal{H}_n$ is a non-null set, since under $\mathcal{H}_n$,
$1/n\leq\upsilon_n\leq (-6R_nK_n\tilde{C}_{\beta}^2+\sqrt{36R_n^2K_n^2\tilde{C}_{\beta}^4+16R_nQ_n\epsilon^2})/(8R_nQ_n)\leq \sqrt{16R_nQ_n\epsilon^2}/(8R_nQ_n)\leq\epsilon/(2(K_n-1)\sqrt{R_n})$, which holds due to Assumption (A).
%with $\upsilon_n$ satisfying $4\upsilon_n^2Q_n+6\upsilon_nK_nn^{\zeta}\leq \epsilon^2$. Then, on $\mathcal{H}_n$,
\begin{align}\label{eq:expression_diff}
&||\bgamma_{n}-\bgamma_{n}^*||_2^2\nonumber\\
&\leq 2R_n\sum_{\bk=(k_1,k_2)\in\mathcal{K}}\left[||\tilde{\bbeta}_{n,k_1}-\tilde{\bbeta}_{n,k_1}^{*}||_2^2\left(2||\tilde{\bbeta}_{n,k_2}-\tilde{\bbeta}_{n,k_2}^*||_2^2+2||\tilde{\bbeta}_{n,k_2}^*||_2^2\right)+||\tilde{\bbeta}_{n,k_2}-\tilde{\bbeta}_{n,k_2}^{*}||_2^2||\tilde{\bbeta}_{n,k_2}^{*}||_2^2\right]\nonumber\\
&\leq 2R_n\sum_{\bk=(k_1,k_2)\in\mathcal{K}}\left[\upsilon_n(2\upsilon_n+2||\tilde{\bbeta}_{n,k_2}^*||_2^2)+\upsilon_n||\tilde{\bbeta}_{n,k_2}^{*}||_2^2\right]
\leq 4Q_nR_n\upsilon_n^2+6\upsilon_nR_nK_n\tilde{C}_{\beta}^2\leq \epsilon^2,
\end{align}
where the last inequality follows from Assumption (D).
Note that, $\Pi(\mathcal{A}_n)\geq\Pi(\mathcal{H}_n)\geq\Pi(\mathcal{H}_n|\mathcal{D}_n)\Pi(\mathcal{D}_n)$ and by Lemma~\ref{lemma_app1}, 
$-\log(\Pi(\mathcal{D}_n)\leq MR_n\log(2+R_n^\omega)$.
From (\ref{eq:expression_diff}),
\begin{align*}
&-\log(\Pi(\mathcal{A}_n)\leq -\log(\Pi(\mathcal{H}_n)\leq -\log(\Pi(\mathcal{D}_n) -\log(\Pi(\mathcal{H}_n|\mathcal{D}_n))\leq -\log(\Pi(\mathcal{H}_n|\mathcal{D}_n))-\log(\Pi(\mathcal{D}_n))\\
%&\leq \sum_{k=1}^{K_n}\sum_{m=1}^M\sum_{r=1}^{R_{n,m}^*}|\beta_{n,m,k}^{*(r)}|^2+\frac{R_nK_nM}{2}\log(2\pi)+\log\left(\frac{R_nK_nM+1}{R_nK_nM}\right)+\frac{R_nK_nM}{2}\log\left(\frac{R_nM}{4}\right)\\
&\leq \sum_{k=1}^{K_n}\sum_{m=1}^M\sum_{r=1}^{R_{n,m}^*}\frac{|\beta_{n,m,k}^{*(r)}|^2}{2}+\frac{R_nK_nM}{2}\log(2\pi)+\log\left(R_nK_nM+1\right)+\frac{R_nK_nM}{2}\log\left(R_nM\right)\\
&\qquad\qquad\qquad\qquad\qquad+R_nK_nM+1+\frac{R_nK_nM}{2}\log\left(\frac{1}{\upsilon_n}\right)+\frac{K_n\upsilon_n}{4}+MR_n\log(2+R_n^\omega)\\
&\leq \frac{\left(\sum_{k=1}^{K_n}\sum_{m=1}^M||\tilde{\bbeta}_{n,m,k}^*||_2\right)^2}{2}+\frac{R_nK_nM}{2}\log(2\pi)+\log\left(R_nK_nM+1\right)+\frac{R_nK_nM}{2}\log\left(R_nM\right)\\
&\qquad\qquad\qquad\qquad\qquad+R_nK_nM+1+R_nK_nM\log\left(\frac{1}{\upsilon_n}\right)+\frac{K_n\upsilon_n}{4}+MR_n\log(2+R_n^\omega),
%&+\frac{R_nK_nM}{4}\log\left(\frac{1}{\upsilon_n}\right)+K_n\upsilon_n+MR_n\log(2+R_n^\omega)\leq \tilde{C}n\epsilon^2, \mbox{\textbf{need upper bound on $\upsilon_n$}}
\end{align*}
where the inequality in the second line follows from the conclusion of Lemma~\ref{lem1} by replacing $\Delta_n^2$ by $\upsilon_n$.
Here $R_nK_n\log(1/\upsilon_n)\leq R_nK_n\log(n)\prec n$ under $\mathcal{H}_n$, by Assumption (A). Also, under $\mathcal{H}_n$, 
$\upsilon_n\leq \epsilon/(2(K_n-1)\sqrt{R_n})\leq \epsilon$, so that $K_n\upsilon_n\prec n$ by Assumption (A). Using Assumptions (A) and (D) to bound all other terms, we arrive at 
$-\log(\Pi(\mathcal{A}_n)\prec n\epsilon^2$, for any $\epsilon>0$. This proves the result.
\end{proof}

\noindent\underline{\emph{Proof of Theorem 3.1}}\\ %~\ref{theorem1}}}\\
To begin with, we define a few metrics of discrepancy between $f(\by_n|x,\bgamma_n)$ and $f(\by_n|x,\bgamma_n^*)$ as below:
\begin{align*}
h_0(f,f^*)&=\int\int f(\by_n|x,\bgamma_n^*)\log\left(\frac{f(\by|x,\bgamma_n^*)}{f(\by|x,\bgamma_n)}\right)\nu_{x}(dx)\nu_{\by_n}(d\by_n),\nonumber\\
h_t(f,f^*)&=(1/t)\left\{\int\int f(\by_n|x,\bgamma_n^* )\left\{\frac{f(\by_n|x,\bgamma_n^*)}{f(\by_n|x,\bgamma_n)}\right\}^t\nu_{\by_n}(d\by_n)\nu_{x}(dx)-1\right\}.
\end{align*}
For every $n$, define a set of probability densities given by $\mathcal{P}_n$. Let the minimum number of Hellinger balls of radius $\epsilon$ required to cover $\mathcal{P}_n$ be given by $\mathcal{N}_{\epsilon}(\mathcal{P}_n)$. To prove the theorem, it suffices
to show that the following conditions (i)-(iii) hold for all large $n$:
%\begin{enumerate}[(i)]
(i) $\log\mathcal{N}_{\epsilon}(\mathcal{P}_n)\leq n\epsilon^2$;
(ii) $\Pi(\mathcal{P}_n^c)\leq \exp(-2n\epsilon^2)$
(iii) $\Pi[f:h_1(f,f^*)\leq\epsilon^2/4]\geq e^{-n\epsilon^2}$,
%\end{enumerate}
using Proposition 1 of \citealp{jiang2007bayesian}. Below we show (i)-(iii) for the proposed model.

\noindent\emph{Proof of condition (i):}  Define $\mathcal{P}_n$ as the set of all densities s.t. at most $s_n$ among $\tilde{\bbeta}_{n,1},...,\tilde{\bbeta}_{n,K_n}$ are nonzero and each element in a nonzero $\tilde{\bbeta}_{n,k}$ satisfies $|\beta_{n,m,k}^{(r)}|\leq C_n$, for $k=1,...,K_n$, where $C_n$ is chosen such that $R_ns_n\log(C_n)\prec n$ and $(1-\Phi(C_n))\leq \exp(-4n\epsilon^2)$, for any $\epsilon>0$.
Such a sequence $C_n$ exists, e.g., $C_n=n$ satisfies the requirement by Assumption (A).
Let $f_{\tilde{\bzeta}}$ denote a density in $\mathcal{P}_n$ expressed with the $K_n$ node-specific inclusion indicators $\tilde{\bzeta}=(\tilde{\zeta}_1,...,\tilde{\zeta}_{K_n})^T$. With $|\tilde{\bzeta}|=\sum_{k=1}^{K_n}\tilde{\zeta}_k$, $\mathcal{P}_n$ contains densities $\tilde{f}_{\tilde{\bzeta}}$ s.t.
$|\tilde{\bzeta}|\leq s_n$.
 Note that, each $\tilde{f}_{\tilde{\bzeta}}\in\mathcal{P}_n$ is represented by $|\tilde{\bzeta}|$ nonzero $\tilde{\bu}_{n,k}$'s with
each component $u_{n,m,k}^{(r)}$, $r=1,...,R_n$, $m=1,...,M$ of a nonzero $\tilde{\bu}_{n,k}$ is bounded between $[-C_n,C_n]$. It takes at most $\left(1+\frac{C_n}{\rho}\right)^{R_nM|\tilde{\bzeta}|}$ balls of the form $[\tilde{\Delta}_{n,m,k}^{(r)}-\rho,\tilde{\Delta}_{n,k}^{(r)}+\rho]$ (with their centers $\tilde{\Delta}_{n,m,k}^{(r)}$'s satisfying $|\tilde{\Delta}_{n,m,k}^{(r)}|\leq C_n$) to cover the parameter space of $\tilde{f}_{\tilde{\bzeta}}$. There are at most $K_n^l$ models
satisfying $|\tilde{\zeta}|=l$. Hence, the total number of balls to cover the parameter space of regression functions in $\mathcal{P}_n$ is given
by $N(\rho)=\sum_{l\leq s_n}K_n^{l}\left(1+\frac{C_n}{\rho}\right)^{R_nMl}\leq (s_n+1)K_n^{s_n}\left(1+\frac{C_n}{\rho}\right)^{MR_ns_n}$.

%Let $\tilde{g}_{\tilde{\bzeta}}$ be any density in $\mathcal{P}_n$, with $\tilde{g}_{\tilde{\bzeta}}=\prod_{m=1}^M\prod_{\bk\in\mathcal{K}}N(y_{n,m,\bk}|xU_{n,m,\bk},1)$ and $U_{n,m,\bk}=\sum_{r=1}^{R_n}\lambda_r u_{n,m,k_1}^{(r)}u_{n,m,k_2}^{(r)}$, where $|u_{n,m,k}^{(r)}|\leq C_n$ and there is exactly $|\tilde{\zeta}|$ nonzero vectors among $\tilde{\bu}_{n,k}$.
%Let $\tilde{f}_{\tilde{\bzeta}}$ be any density in $\mathcal{P}_n$, with
%$\tilde{f}(\bY|x)=\prod_{\bj\in\mathcal{J}}\tilde{f}_{\bj}(Y_{\bj}|x),\:\tilde{f}_{\bj}(Y_{\bj}|x)=\exp(a(\tilde{\alpha}_{\bj})Y_{\bj}+b(\tilde{\alpha}_{\bj})+c(Y_{\bj}))$, $\tilde{\alpha}_{\bj}=x\tilde{B}_{\bj}$,
%where $\tilde{B}_{\bj}=\sum_{r=1}^{R_n} v_{j_1}^{(r)}...v_{j_D}^{(r)}$, with $|v_{k}^{(r)}|\leq C_n$ for all $k\in\mathcal{N}$, $r=1,..,R_n$. There exists
%a density $f\in\mathcal{P}_n$ represented by parameters $u_k^{(r)}$'s such that $v_k^{(r)}\in(u_k^{(r)}-\rho,u_k^{(r)}+\rho)$ for every $r$ and $k$. 
%Now note that,
%\begin{align*}
%&|u_{n,m,\bk}-v_{n,m,\bk}|=|\sum_{r=1}^{R_n}\lambda^{(r)} u_{k_1}^{(r)}u_{k_2}^{(r)}-\sum_{r=1}^{R_n}\lambda^{(r)} v_{k_1}^{(r)}v_{k_2}^{(r)}|
%\leq \sum_{r=1}^{R_n}\Big\{|u_{k_1}^{(r)}-v_{k_1}^{(r)}| |u_{k_2}^{(r)}|+
%|v_{k_1}^{(r)}||u_{k_2}^{(r)}-v_{k_2}^{(r)}\Big\}\\
%&\leq R_n\rho C_n.
%\end{align*}
Let $\tilde{f}_{\tilde{\bzeta}}$ be a density in $\mathcal{P}_n$ represented by $|\tilde{\bzeta}|$ nonzero $\tilde{\bu}_{n,k}$'s with
$u_{n,m,k}^{(r)}$, $r=1,...,R_n$, $m=1,...,M$ as components of a nonzero $\tilde{\bu}_{n,k}$. There exist a density $\tilde{\tilde{f}}_{\tilde{\bzeta}}$ in $\mathcal{P}_n$ represented by $|\tilde{\bzeta}|$ nonzero $\tilde{\bv}_{n,k}$'s with each component $v_{n,m,k}^{(r)}$, $r=1,...,R_n$, $m=1,...,M$ of a nonzero $\tilde{\bv}_{n,k}$ satisfies 
$u_{n,m,k}^{(r)}\in (v_{n,m,k}^{(r)}-\rho,v_{n,m,k}^{(r)}+\rho)$.

Note that, $h(\tilde{\tilde{f}},\tilde{f})\leq \left\{\sum_{m=1}^{M}\sum_{\bk\in\mathcal{K}}h_0(\tilde{\tilde{f}}_{m,\bk},\tilde{f}_{m,\bk})\right\}^{1/2}$. Applying Taylor expansion, we have 
$h_0(\tilde{\tilde{f}}_{m,\bk},\tilde{f}_{m,\bk})\leq E_x[\{a_m'(\bar{H}_{m,\bk})(-b_m'(\bar{H}_{m,\bk})/a_m'(\bar{H}_{m,\bk}))+b_m'(\bar{H}_{m,\bk})\}(x\tilde{\tilde{\gamma}}_{n,m,\bk}-x\tilde{\gamma}_{n,m,\bk})]$, where $\bar{H}_{m,\bk}$ is an intermediate point between $x\tilde{\tilde{\gamma}}_{n,m,\bk}$ and $x\tilde{\gamma}_{n,m,\bk}$,  $\tilde{\gamma}_{n,m,\bk}=\sum_{r=1}^{R_{n}}\lambda_{n,m}^{(r)} u_{n,m,k_1}^{(r)}u_{n,m,k_2}^{(r)}$ and $\tilde{\tilde{\gamma}}_{n,m,\bk}=\sum_{r=1}^{R_{n}}\lambda_{n,m}^{(r)} v_{n,m,k_1}^{(r)}v_{n,m,k_2}^{(r)}$. Now note that,
\begin{align*}
&|\tilde{\tilde{\gamma}}_{n,m,\bk}-\tilde{\gamma}_{n,m,\bk}|=|\sum_{r=1}^{R_{n}}\lambda_{n,m}^{(r)} u_{n,m,k_1}^{(r)}u_{n,m,k_2}^{(r)}-\sum_{r=1}^{R_n}\lambda_{n,m}^{(r)} v_{n,m,k_1}^{(r)}v_{n,m,k_2}^{(r)}|\nonumber\\
&\leq \sum_{r=1}^{R_n}\Big\{|u_{n,m,k_1}^{(r)}-v_{n,m,k_1}^{(r)}| |u_{k_2}^{(r)}|+
|v_{k_1}^{(r)}||u_{k_2}^{(r)}-v_{k_2}^{(r)}\Big\}
\leq R_n\rho C_n.
\end{align*}
Following the similar arguments,
$|\bar{H}_{m,k}|\leq\max\{|x\tilde{\gamma}_{n,m,\bk}|,|x\tilde{\tilde{\gamma}}_{n,m,\bk}|\}\leq R_nC_n^2$,
\begin{align}
h(\tilde{\tilde{f}},\tilde{f})\leq \left\{2\rho Q_nMR_nC_n\max_{m=1:M}\sup_{|w|\leq R_nC_n^2}|a_m'(w)|\sup_{|w|\leq R_nC_n^2}|b_m'(w)/a_m'(w)|\right\}^{1/2}.
\end{align}
Thus choosing
$\rho=\epsilon^2/[R_nC_nMQ_n\max_{m=1:M}\{\sup_{|w|\leq R_nC_n^2}|a_m'(w)|\sup_{|w|\leq R_nC_n^2}|b_m'(w)/a_m'(w)|\}]$, we have $h(\tilde{\tilde{f}},\tilde{f})\leq \epsilon$. Hence,
%Similar calculations lead to $|\alpha_{\bj}|$, $|\tilde{\alpha}_{\bj}|$ (and therefore $|\bar{\alpha}_{\bj}|$) being bounded by $R_nC_n$. Hence,
%\begin{align*}
%\mathcal{D}_H(f,\tilde{f})\leq \left\{\sum_{\bj\in\mathcal{J}}\mathcal{D}_0(f_{\bj},\tilde{f}_{\bj})\right\}^{1/2}
%\leq \{2q_n \sup\limits_{|w|\leq R_nC_n^D}|a'(w)|\sup\limits_{|w|\leq R_nC_n^D}|b'(w)/a'(w)|\rho R_nC_n^{D-1}\}^{1/2}.
%\end{align*}
%Choosing $\rho=\epsilon_n^2/(2q_n \sup\limits_{|w|\leq R_nC_n^D}|a'(w)|\sup\limits_{|w|\leq R_nC_n^D}|b'(w)/a'(w)| R_nC_n^{D-1})$, one gets
%$\mathcal{D}_H(f,\tilde{f})\leq \epsilon_n$. Hence
\begin{align*}
&\log\mathcal{N}_{\epsilon}(\mathcal{P}_n)\leq \log N(\rho)\\
&=MR_ns_n\log\left(1+\frac{MQ_nR_nC_n^2}{\epsilon^2}\max_{m=1:M}\left\{\sup_{|w|\leq R_nC_n^2}|a_m'(w)|\sup_{|w|\leq R_nC_n^2}|b_m'(w)/a_m'(w)|\right\}\right)\\
&\leq R_{n}s_{n}M\log\left(Q_{n}M/\epsilon^2\right)+R_ns_n\log(F(R_nC_n^2))+s_n\log(K_n)+\log(s_n+1)\\
%&\leq R_ns_nM\log(C_n/\epsilon_n)+R_ns_nM\log(\sqrt{Q_nM}R_n+1)+
%s_n\log(K_n)+\log(s_n+1)\\
%&\leq DR_np_n \log(2p_n)+R_np_n\log(1/\epsilon_n^2)+R_np_n\log(G(R_nC_n^D))\\
&\leq n\epsilon^2,\:\:\mbox{for large $n$, by Assumptions (A) and (E).}
\end{align*}\

\noindent\emph{Proof of condition (ii):} 
%Define $\mathcal{P}_n$ as the set of all densities s.t. at most $m_n$ among $\tilde{\bgamma}_{1},...,\tilde{\bgamma}_{V_n}$ are nonzero and each element in a nonzero $\tilde{\bgamma}_{h}$ satisfies $|\gamma_{h}^{(r)}|\leq C_n$, for $h=1,...,V_n$. Let $g_{\bzeta}$ denote a density in $\mathcal{P}_n$ expressed with the binary variables $\bzeta=(\zeta_1,...,\zeta_{V_n})'$. With $|\bzeta|=\sum_{h=1}^{V_n}\zeta_h$, $\mathcal{P}_n$ contains densities $g_{\bzeta}$ s.t.
%$|\bzeta|\leq m_n$. 
As defined in condition (i), $\mathcal{A}=\{k\in\mathcal{N}:\tilde{\zeta}_k=1\}$. Then for all large $n$,
\begin{align*}
\Pi(\mathcal{P}_n^c)&=%\Pi(|\bzeta|>m_n)+
\sum_{|\bzeta|\leq s_n}\Pi(\cup_{k\in\mathcal{A}}\cup_{m=1}^M\cup_{r=1}^{R_n}\{|\beta_{n,m,k}^{(r)}|>C_n\})\Pi(\tilde{\bzeta})
\leq \max_{\tilde{\bzeta}:|\tilde{\bzeta}|\leq s_n}\Pi(\cup_{k\in\mathcal{A}}\cup_{m=1}^M\cup_{r=1}^{R_n}\{|\beta_{n,m,k}^{(r)}|>C_n\})\\
&\leq R_ns_nM\Pi(|\beta_{n,m,k}^{(r)}|>C_n)
=2R_ns_nM(1-\Phi(C_n))
\leq\exp(\log(2R_ns_nM))(1-\Phi(C_n))\\
&\leq exp(-2n\epsilon^2),
\end{align*}
for all large $n$,
where the last inequality follows from Assumption (A) and by the choice of $C_n$. \\

\emph{Proof of Condition (iii):} Consider $t=1$. By mean value theorem, $\exists\:\bkappa$ s.t. $h_1(f,f^*)=E_x\left\{{\boldsymbol f}'(\bkappa)^T(x\bgamma_n-x\bgamma_n^*)\right\}$, where ${\boldsymbol f}'$ represents the continuous derivative function of $f$ in the neighborhood of $f^*$. Let $\delta_n=\epsilon^2/(2a_0MQ_n)$. If for each $\bk\in\mathcal{K}$, $\gamma_{n,m,\bk}\in(\gamma_{n,m,\bk}^*-\delta_n,\gamma_{n,m,\bk}^*+\delta_n)$, then
$||x\bgamma_{n}-x\bgamma_{n}^*||\leq  \sum_{m=1}^M\sum_{\bk\in\mathcal{K}}|x\gamma_{n,m,\bk}-x\gamma_{n,m,\bk}^*|\leq Q_nM\delta_n\leq \epsilon^2$.
Again, $||\bkappa||\leq a_0||\bgamma_n-\bgamma_n^*||+a_0||\bgamma_n^*||\leq \epsilon^2+a_0||\bgamma_n^*||$, where $||\bgamma_n^*||\leq \sum_{m=1}^M||\bgamma_{n,m}^*||\leq\sum_{m=1}^M\sum_{\bk\in\mathcal{K}}||\tilde{\bbeta}_{n,k_1}^*|| ||\tilde{\bbeta}_{n,k_2}^*||\leq \sum_{m=1}^M(\sum_{k=1}^{K_n}||\tilde{\bbeta}_{n,k}^*||)^2\leq (\sum_{m=1}^M\sum_{k=1}^{K_n}||\tilde{\bbeta}_{n,k}^*||)^2$, which is bounded by Assumption (D), for sufficiently large $n$. Hence $||{\boldsymbol f}'(\bkappa)||$ is bounded for sufficiently large $n$.
Thus, $h_t(f,f^*)=E_x\left\{{\boldsymbol f}_1(\bkappa)^T(x\bgamma_n-x\bgamma_n^*)\right\}\leq \tilde{C} Q_nM\delta_n\leq \epsilon^2/4$ for large $n$, for some constant $\tilde{C}>0$.

This implies that $\Pi(\{f:h_t(f,f^*)\leq\epsilon^2/4\})\geq \Pi(\{\bgamma_{n,\bk}:\gamma_{n,m,\bk}\in(\gamma_{n,m,\bk}^*-\delta_n,\gamma_{n,m,\bk}^*+\delta_n),\forall\:\bk\in\mathcal{K}\})$.
By the calculations in Lemma~\ref{lem2},
$-\log \Pi(\{\bgamma_{n}:\gamma_{n,m,\bk}\in(\gamma_{n,m,\bk}^*-\delta_n,\gamma_{n,m,\bk}^*+\delta_n),\forall\:\bk\in\mathcal{K}\})=-\log\Pi(||\bgamma_{n}-\bgamma_n^*||_{\infty}\leq \delta_n)\leq
\frac{\left(\sum_{k=1}^{K_n}\sum_{m=1}^M||\tilde{\bbeta}_{n,m,k}^*||_2\right)^2}{2}+\frac{R_nK_nM}{2}\log(2\pi)+\log\left(R_nK_nM+1\right)+\frac{R_nK_nM}{2}\log\left(R_nM\right)+R_nK_nM+1+R_nK_nM\log\left(\frac{1}{\tilde{\upsilon}_n}\right)+\frac{K_n\tilde{\upsilon_n}}{4}+MR_n\log(2+R_n^\omega)$, where $1/n\leq \tilde{\upsilon}_n\leq \epsilon/(2(K_n-1)\sqrt{2a_0MQ_nR_n})$. 
Thus, by Assumptions (A) and (D), 
$-\log\Pi(||\bgamma_{n}-\bgamma_n^*||_{\infty}\leq \delta_n)\prec n\epsilon^2$ for any $\epsilon>0$, proving condition (iii).

\noindent\textbf{Proof of Theorem 3.2}\\ %~\ref{theorem2}}\\
Define, $\mathcal{B}_n=\left\{\by_n:\int \frac{f(\by_n|x,\bgamma_n)}{f(\tilde{\by}_n|x,\bgamma_n^*)} \pi(\bgamma_n)d\bgamma_n\geq \exp(- n\epsilon^2)\right\}$ as in Lemma~\ref{lem2}, and let $\mathcal{C}_n=\{|\tilde{\zeta}|>C_0s_n\}$, for any large constant $C_0$. Now note that,
\begin{align}\label{eq:set_nonzero}
&\Pi(\mathcal{C}_n)=\sum_{k=\floor{s_nc_0}+1}^{K_n}{K_n\choose k}\int \eta^k(1-\eta)^{K_n-k}\pi(\eta)d\eta=\sum_{k=\floor{s_nc_0}+1}^{K_n}{K_n\choose k} Beta(k+1,K_n-k+b_{\eta})\nonumber\\
&=\sum_{k=\floor{s_nC_0}+1}^{K_n} \frac{K_n!}{k!(K_n-k)!}\frac{(K_n-k+b_{\eta}-1)!k!}{(K_n+b_{\eta})!}\leq\prod_{k=1}^{b_{\eta}}\left(1-\frac{\floor{s_nC_0}+2}{K_n+k}\right)\leq\left(1-\frac{\floor{s_nC_0}+2}{K_n+1}\right)^{b_{\eta}}\nonumber\\
&\leq \exp\left(-b_{\eta}(\floor{s_nC_0}+2)/(K_n+1)\right)\leq \exp(-2n\epsilon^2),\:\mbox{by Assumption (F)}.
\end{align}
Thus,
\begin{align}\label{eq_initial}
& E_{\bgamma_n^*}\Pi(\mathcal{C}_n|\by_n,x_1,...,x_n)\leq E_{\bgamma_n^*}\left[\frac{\int_{\mathcal{C}_n}\frac{f(\by_n|x,\bgamma_n)}{f(\by_n|x,\bgamma_n^*)}\pi(\bgamma_n)d\bgamma_n}{\int \frac{f(\by_n|x,\bgamma_n)}{f(\by_n|x,\bgamma_n^*)}\pi(\bgamma_n)d\bgamma_n}1_{\by_n\in\mathcal{B}_n}\right]+P_{\bgamma_n^*}(\mathcal{B}_n^c)\nonumber\\
&\leq E_{\bgamma_n^*}\left[\int_{\mathcal{C}_n}\frac{f(\by_n|x,\bgamma_n)}{f(\by_n|x,\bgamma_n^*)}\pi(\bgamma_n)d\bgamma_n1_{\by_n\in\mathcal{B}_n}\right]\exp(n\epsilon^2)+P_{\bgamma_n^*}(\mathcal{B}_n^c)\nonumber\\
&\leq\Pi(\mathcal{C}_n)\exp(n\epsilon^2)+P_{\bgamma_n^*}(\mathcal{B}_n^c)\leq \exp(-n\epsilon^2)+P_{\bgamma_n^*}(\mathcal{B}_n^c)\rightarrow 0
\end{align}
where the last line follows from the conclusions of Lemma~\ref{lem2} and equation (\ref{eq:set_nonzero}).

\section*{Appendix B}

Let $q$ be the cardinality of the upper-triangular edge index set $\mathcal{K}$. The full conditional distributions for the model parameters are given by 

\begin{enumerate}
    \item $\mu_{1}|- \sim N(\frac{1}{\sigma_{1}^{2} + nq} \sum_{i=1}^{n}\sum_{\underline{k} \in \mathcal{K}} y_{i, 1, \underline{k}} - x_i \gamma_{1,\underline{k}} - \tilde{x}_i \tilde{\alpha}_1 , \; \frac{\sigma_{1}^{2}}{\sigma_{1}^{2} + nq})$,
    
    where $\gamma_{1,\underline{k}}$ is the edge coefficient on layer 1 indexed by $\underline{k}$.

    \item $\mu_{2}|- \sim N(\frac{1}{\sigma_{2}^{2} + nq} \sum_{i=1}^{n}\sum_{\underline{k} \in \mathcal{K}} y_{i, 2, \underline{k}} - x_i \gamma_{2,\underline{k}} - \tilde{x}_i \tilde{\alpha}_2 , \; \frac{\sigma_{2}^{2}}{\sigma_{2}^{2} + nq})$,
    
    where $\gamma_{2,\underline{k}}$ is the edge coefficient on layer 2 indexed by $\underline{k}$.

    \item $\sigma_1^2 | - \sim IG(a + \frac{nq}{2}, \; b + \frac{1}{2} \sum_{i=1}^{n}\sum_{\underline{k} \in \mathcal{K}} (y_{i, 1, \underline{k}} - \mu_1 - x_i \gamma_{1,\underline{k}} - \tilde{x}_i \tilde{\alpha}_1)^2 )$.

    \item $\sigma_2^2 | - \sim IG(a + \frac{nq}{2}, \; b + \frac{1}{2} \sum_{i=1}^{n}\sum_{\underline{k} \in \mathcal{K}} (y_{i, 2, \underline{k}} - \mu_2 -  x_i \gamma_{2,\underline{k}} - \tilde{x}_i \tilde{\alpha}_2)^2 )$.

    \item $\tilde{\alpha}_1 | - \sim N(\frac{1}{\sigma_{1}^{2} + q\sum_{i=1}^{n} \tilde{x}_{i}^{2}}  \sum_{i=1}^{n}\sum_{\underline{k} \in \mathcal{K}} \tilde{x}_i z_{i,1,\underline{k}}, \; \frac{\sigma_{1}^{2}}{\sigma_{1}^{2} + q\sum_{i=1}^{n} \tilde{x}_{i}^{2}})$,
    
    where $z_{i,1,\underline{k}} = y_{i, 1, \underline{k}} - \mu_1 - x_i \gamma_{1,\underline{k}}$.

    \item $\tilde{\alpha}_2 | - \sim N(\frac{1}{\sigma_{2}^{2} + q\sum_{i=1}^{n} \tilde{x}_{i}^{2}}  \sum_{i=1}^{n}\sum_{\underline{k} \in \mathcal{K}} \tilde{x}_i z_{i,2,\underline{k}}, \; \frac{\sigma_{2}^{2}}{\sigma_{2}^{2} + q\sum_{i=1}^{n} \tilde{x}_{i}^{2}})$,
    
    where $z_{i,2,\underline{k}} = y_{i, 2, \underline{k}} - \mu_2 - x_i \gamma_{2,\underline{k}}$.

    \item For the update of $\underline{\bbeta}_k = (\underline{\bbeta}^T_{1,k}, \underline{\bbeta}^T_{2,k})^T$, define: 

    $
    \underline{U}_1 = (\underline{\bbeta}_{1,1}, ..., \underline{\bbeta}_{1,k-1}, \underline{\bbeta}_{1,k+1}, ..., \underline{\bbeta}_{1,K})^T
    $, $
    \underline{U}_2 = (\underline{\bbeta}_{2,1}, ..., \underline{\bbeta}_{2,k-1}, \underline{\bbeta}_{2,k+1}, ..., \underline{\bbeta}_{2,K})^T
    $, 
    
    $
    \underline{y}_{i,1} = (y_{i,1, (1, k)}, ..., y_{i,1, (k-1, k)}, y_{i,1, (k, k+1)}, ..., y_{i,1, (k, K)})^T
    $, and
    
    $
    \underline{y}_{i,2} = (y_{i,2, (1, k)}, ..., y_{i,2, (k-1, k)}, y_{i,2, (k, k+1)}, ..., y_{i,2, (k, K)})^T
    $. 

    Furthermore, let: 

    $\underline{y}_i = \begin{pmatrix} 
    \underline{y}_{i,1} \\
    \underline{y}_{i,2}
    \end{pmatrix}$, $\underline{\mu} = \begin{pmatrix} 
    \mu_1 \mathbf{1}_{K-1} \\
    \mu_2 \mathbf{1}_{K-1}
    \end{pmatrix}$, 
    $\underline{U}_i = \begin{pmatrix}
         x_i \underline{U}_1 \bLambda_1  &  \bzero_{(V-1) \times R} \\ 
       \bzero_{(V-1) \times R} &  x_i \underline{U}_2 \bLambda_2
    \end{pmatrix}$, and
    $\underline{\tilde{x}}_i = \begin{pmatrix} 
    \tilde{x}_i \tilde{\alpha}_1 \mathbf{1}_{K-1} \\
    \tilde{x}_i \tilde{\alpha}_2 \mathbf{1}_{K-1}
    \end{pmatrix}$.

    Lastly, define: 

    $\underline{\tilde{y}}  = \begin{pmatrix}
        \underline{y}_1 \\
        \vdots \\
        \underline{y}_n
    \end{pmatrix}$,
    $\;\underline{\tilde{\mu}} = \mathbf{1}_n \otimes \underline{\mu}$,  
    $\;\underline{\tilde{U}} = \begin{pmatrix}
        \underline{U}_1 \\
        \vdots \\
        \underline{U}_n
    \end{pmatrix}$, $\; \underline{\tilde{x}} = \begin{pmatrix}
        \underline{\tilde{x}}_1 \\
        \vdots \\
        \underline{\tilde{x}}_n
    \end{pmatrix}$,
    $\mathbf{\tilde{A}} = \bI_n \otimes \begin{pmatrix}
        \sigma_1^2 \bI_{K-1} & \bzero_{(K-1) \times (K-1)} \\
        \bzero_{(K-1) \times (K-1)} & \sigma_2^2 \bI_{K-1}
    \end{pmatrix}$,
    
    and $\underline{\tilde{z}} = \underline{\tilde{y}} -  \underline{\tilde{\mu}} - \underline{\tilde{x}}$.

    Then the update for $\underline{\bbeta}_k$ is: 

    $$ 
    \underline{\bbeta}_k |- \sim \xi_k N(\bmu_{\underline{\bbeta}_k}, \bSigma_{\underline{\bbeta}_k}) + (1 - \xi_k) \delta_{\bzero},
    $$

    where $\bSigma_{\underline{\bbeta}_k} = (\bJ^{-1} + \underline{\tilde{U}}^T \mathbf{\tilde{A}}^{-1} \underline{\tilde{U}})^{-1}$ and $\bmu_{\underline{\bbeta}_k} = \bSigma_{\underline{\bbeta}_k} \underline{\tilde{U}}^T \mathbf{\tilde{A}}^{-1} \underline{\tilde{z}}$.

    \item $\xi_k | - \sim Ber(\pi_k)$, where $\pi_k = \frac{\eta N(\underline{\tilde{z}} | \bzero_{2n(K-1)}, \mathbf{\tilde{A}} + \underline{\tilde{U}} \bJ \underline{\tilde{U}}^T)}{\eta N(\underline{\tilde{z}} | \bzero_{2n(K-1)}, \mathbf{\tilde{A}} + \underline{\tilde{U}} \bJ \underline{\tilde{U}}^T) + (1-\eta) N(\underline{\tilde{z}} | \bzero_{2n(K-1)}, \mathbf{\tilde{A}})} $.

    \item $\lambda_1^{(r)} | - = \begin{cases} 
      0 & w.p. \; p^{(r)}_{1,1}, \\
      1 & w.p. \; p^{(r)}_{1,2}, \\
      -1 & w.p. \; p^{(r)}_{1,3}, 
   \end{cases}$

   where $p^{(r)}_{1,1} = \frac{\pi_{1,1}^{(r)} \prod_{i=1}^{n} N(\by_{i,1}| \mathbf{1}_q \mu_1 + \underline{\bgamma}_1^{(\lambda_1^{(r)} = 0)} x_i + \mathbf{1}_q \tilde{x}_i \tilde{\alpha}_1, \; \sigma_1^2 \bI_{q})}{S}$, 

$p^{(r)}_{1,2} = \frac{\pi_{1,2}^{(r)} \prod_{i=1}^{n} N(\by_{i,1}| \mathbf{1}_q \mu_1 + \underline{\bgamma}_1^{(\lambda_1^{(r)} = 1)} x_i + \mathbf{1}_q \tilde{x}_i \tilde{\alpha}_1, \; \sigma_1^2 \bI_{q})}{S}$, 
   
   $p^{(r)}_{1,3} = \frac{\pi_{1,3}^{(r)} \prod_{i=1}^{n} N(\by_{i,1}| \mathbf{1}_q \mu_1 + \underline{\bgamma}_1^{(\lambda_1^{(r)} = -1)} x_i + \mathbf{1}_q \tilde{x}_i \tilde{\alpha}_1, \; \sigma_1^2 \bI_{q})}{S}$, 
   
   $\by_{i,1} = (y_{i,1,\underline{k}}: \underline{k} \in \mathcal{K})$, and

       $S = \pi_{1,1}^{(r)} \prod_{i=1}^{n} N(\by_{i,1}| \mathbf{1}_q \mu_1 + \underline{\bgamma}_1^{(\lambda_1^{(r)} = 0)} x_i + \mathbf{1}_q \tilde{x}_i \tilde{\alpha}_1, \; \sigma_1^2 \bI_{q}) + $ 

       $\pi_{1,2}^{(r)} \prod_{i=1}^{n} N(\by_{i,1}| \mathbf{1}_q \mu_1 + 
       \underline{\bgamma}_1^{(\lambda_1^{(r)} = 1)} x_i + \mathbf{1}_q \tilde{x}_i \tilde{\alpha}_1, \; \sigma_1^2 \bI_{q}) + $

   $\pi_{1,3}^{(r)} \prod_{i=1}^{n} N(\by_{i,1}| \mathbf{1}_q \mu_1 + \underline{\bgamma}_1^{(\lambda_1^{(r)} = -1)} x_i + \mathbf{1}_q \tilde{x}_i \tilde{\alpha}_1, \; \sigma_1^2 \bI_{q})$. 
   Here, if 
   $\bB_1 = \begin{psmallmatrix}
   \bbeta_{1,1}^T \\
   \vdots \\
   \bbeta_{1,K}^T
   \end{psmallmatrix}$, then $\underline{\bgamma}_1$ is the vector of upper-triangular entries of the network coefficient matrix, $\bB_1 \bLambda_1 \bB_1^T$. Furthermore, $\underline{\bgamma}_1^{(\lambda_1^{(r)} = s)}$, is the resulting vector, $\underline{\bgamma}_1$, when $\lambda_1^{(r)}$ equals $s$, in $\bLambda_1$.

   \item $\lambda_2^{(r)} | - = \begin{cases} 
      0 & w.p. \; p^{(r)}_{2,1}, \\
      1 & w.p. \; p^{(r)}_{2,2}, \\
      -1 & w.p. \; p^{(r)}_{2,3},
   \end{cases}$

   where $p^{(r)}_{2,1}$, $p^{(r)}_{2,2}$, and $p^{(r)}_{2,3}$ are defined analogously to $p^{(r)}_{1,1}$, $p^{(r)}_{1,2}$, and $p^{(r)}_{1,3}$, respectively.

   \item $\bJ | - \sim IW(\nu + \#\{k: \xi_k = 1\}, \bI_{2R} + \sum_{\{k: \xi_k = 1\}} \underline{\bbeta}_k \underline{\bbeta}_k^T)$.

   \item $(\pi^{(r)}_{1,1}, \pi^{(r)}_{1,2}, \pi^{(r)}_{1,3}) | - \sim Dirichlet(r^\omega + I(\lambda_1^{(r)} = 0), 1 + I(\lambda_1^{(r)} = 1), 1 + I(\lambda_1^{(r)} = -1))$.

   \item $(\pi^{(r)}_{2,1}, \pi^{(r)}_{2,2}, \pi^{(r)}_{2,3}) | - \sim Dirichlet(r^\omega + I(\lambda_2^{(r)} = 0), 1 + I(\lambda_2^{(r)} = 1), 1 + I(\lambda_2^{(r)} = -1))$.

   \item $\eta | - \sim Beta(a + \#\{k: \xi_k = 1\}, b + K - \#\{k: \xi_k = 1\})$. 
    
\end{enumerate}

\section*{Appendix C}
\begin{table}[H]
    \centering
    \setlength{\extrarowheight}{-9pt}
    \begin{tabular}{|c|c|c|c|}
    \hline
    \multicolumn{2}{|c|}{ROI Names} \\
 \hline
LH.VisCent.Striate.1  &   LH.VisCent.ExStr.4 \\ 
LH.VisCent.ExStr.5   &    LH.VisPeri.ExStrInf.2 \\  
 LH.VisPeri.ExStrInf.3 &  LH.SomMotA.1 \\     
 LH.SomMotA.2  &           LH.SomMotA.3  \\          
 LH.SomMotA.8 &            LH.SomMotB.Aud.1  \\    
 LH.SomMotB.Cent.2 &       LH.DorsAttnA.TempOcc.2 \\ 
LH.DorsAttnA.SPL.1  &     LH.DorsAttnB.PostC.2 \\   
LH.DorsAttnB.PostC.4  &   LH.DorsAttnB.FEF.1   \\   
LH.SalVentAttnA.FrMed.1  & LH.SalVentAttnA.FrMed.2 \\ 
LH.SalVentAttnB.IPL.1 &   LH.SalVentAttnB.PFCl.1 \\ 
LH.LimbicB.OFC.1 &        LH.LimbicB.OFC.2   \\   
LH.LimbicA.TempPole.2  &  LH.LimbicA.TempPole.3  \\ 
LH.LimbicA.TempPole.4&    LH.ContA.IPS.1   \\    
LH.ContA.IPS.2   &        LH.ContA.PFCd.1 \\        
LH.ContA.PFCl.1 &         LH.ContB.Temp.1 \\    
LH.ContB.IPL.1  &         LH.ContB.PFClv.1 \\       
LH.ContC.pCun.1 &         LH.ContC.Cingp.1   \\ 
LH.DefaultA.pCunPCC.1 &   LH.DefaultA.pCunPCC.3 \\  
LH.DefaultA.PFCm.3 &      LH.DefaultB.Temp.2  \\ 
LH.DefaultC.IPL.1 &       LH.DefaultC.Rsp.1  \\     
LH.TempPar.1 &            LH.TempPar.2      \\   
RH.VisCent.Striate.1 &    RH.VisCent.ExStr.3  \\    
RH.VisCent.ExStr.4 &      RH.SomMotA.1 \\       
RH.SomMotA.4  &           RH.SomMotA.7     \\       
 RH.SomMotA.8  &           RH.SomMotB.Aud.2 \\   
 RH.SomMotB.S2.1 &         RH.DorsAttnA.TempOcc.1 \\ 
 RH.DorsAttnA.ParOcc.1  &  RH.DorsAttnA.SPL.3 \\   
 RH.DorsAttnB.PostC.1 &    RH.DorsAttnB.PostC.3  \\  
 RH.DorsAttnB.PostC.4 &    RH.DorsAttnB.FEF.1  \\  
 RH.SalVentAttnA.Ins.1 &   RH.SalVentAttnA.FrOper.1 \\
 RH.SalVentAttnA.ParMed.1 & RH.SalVentAttnA.ParMed.2  \\
 RH.SalVentAttnB.IPL.1 &    RH.SalVentAttnB.PFClv.1 \\ 
 RH.SalVentAttnB.PFCl.1 &   RH.SalVentAttnB.Ins.1 \\ 
 RH.SalVentAttnB.Ins.2 &    RH.SalVentAttnB.PFCmp.1 \\
 RH.LimbicB.OFC.3 &        RH.LimbicB.OFC.4  \\ 
 RH.ContA.IPS.1 &          RH.ContA.IPS.2  \\        
 RH.ContA.PFCl.2 &         RH.ContB.Temp.2 \\ 
 RH.ContB.IPL.1 &          RH.ContB.PFCld.1 \\       
 RH.ContB.PFClv.1  &       RH.ContB.PFCmp.1 \\ 
 RH.ContB.PFCld.3 &        RH.ContC.pCun.2 \\        
 RH.ContC.Cingp.1  &       RH.DefaultA.IPL.1  \\ 
 RH.DefaultA.pCunPCC.1 &   RH.DefaultA.PFCm.1 \\      
 RH.DefaultB.PFCv.1&       RH.DefaultC.IPL.1 \\ 
 RH.DefaultC.Rsp.1 &       RH.TempPar.4 \\
 \hline
    \end{tabular}
    \caption{Names of the ROIs selected by JL as being associated with the MMSE outcome.}
    \label{Selected_ROI_table}
\end{table}

\bibliographystyle{natbib}
\bibliography{References}